\documentclass[10pt,journal,compsoc,twoside]{IEEEtran}
\usepackage[british]{babel}

\usepackage{tikz}%
\usetikzlibrary{matrix,positioning,arrows,fit,shapes.multipart,%
  decorations.pathreplacing,calc}%

\tikzset{%
  table/.style={%
    matrix of nodes, %
    row sep=-\pgflinewidth, %
    column sep=-\pgflinewidth, %
    nodes={rectangle,draw=black,text width=#1,align=center}, %
    text depth=.5ex, %
    text height=2ex, %
    nodes in empty cells %
  }%
}

\usepackage{etoolbox}
\newtoggle{draft}

\iftoggle{draft}{%
  \usepackage{todonotes}
}{%
  \usepackage[disable]{todonotes}
}%

\let\labelindent\relax
\usepackage[shortlabels,inline]{enumitem}
\setlist{labelindent=0pt, leftmargin=*, nosep}
\SetEnumitemKey{iff}{wide, font=\normalfont\itshape}
\SetEnumitemKey{claim}{label=Claim \arabic*., ref=Claim \arabic*, %
  itemsep=3pt plus 1pt minus 1pt, partopsep=3pt plus 1pt minus 1pt, %
  font=\normalfont\bfseries, wide, leftmargin=\parindent}
\SetEnumitemKey{claimproof}{leftmargin=\parindent, labelindent=\parindent, %
    labelwidth=*, partopsep=3pt plus 1pt minus 1pt}

\usepackage{tabularx,booktabs,multirow}

\makeatletter 
\newcommand\mynobreakpar{\par\nobreak\@afterheading} 
\makeatother

\ifCLASSOPTIONcompsoc
\else
\fi
\ifCLASSINFOpdf
\else
\fi
\usepackage[cmex10]{amsmath}
\usepackage{amssymb}
\usepackage{amsthm}
\usepackage{thmtools}

\newtheorem{definition}{Definition}
\newtheorem{theorem}{Theorem}
\newtheorem{lemma}{Lemma}

\newtheoremstyle{proof}%
{}
{1ex}
{}
{}
{\bf}
{.}
{.5em}
{}

\declaretheoremstyle[%
notefont=\bfseries,%
notebraces={}{},%
bodyfont=\normalfont,%
headformat=Proof\NOTE%
]{nopar}
\declaretheorem[style=nopar]{thmpf}

\theoremstyle{proof}
\newtheorem*{pf}{Proof}

\newtheorem{example}{Example}

\usepackage{url}
\makeatletter
\def\url@stt{\def\UrlFont{\small\ttfamily}}
\makeatother
\usepackage{hyperref}
\hypersetup{%
  colorlinks,
  linkcolor={red},
  citecolor={blue},
  urlcolor={blue!90!black}
}

\DeclareMathOperator{\var}{\sf var}
\DeclareMathOperator{\sig}{\sf sig}
\DeclareMathOperator{\const}{\sf const}

\renewcommand{\to}{\ensuremath{\rightarrow}}
\renewcommand{\emptyset}{\varnothing}

\newcommand{\sign}[1]{\ensuremath{\mathbf{#1}}}
\newcommand{\lang}[1]{\ensuremath{\mathfrak{#1}}}
\newcommand{\arity}[1]{\ensuremath{\left|#1\right|}}
\newcommand{\R}{\ensuremath{\sign{R}}}
\newcommand{\V}{\ensuremath{\sign{V}}}
\newcommand{\tuple}[1]{\ensuremath{\langle#1\rangle}}

\newcommand{\dom}{\ensuremath{\mathbf{dom}}}
\newcommand{\adom}{\ensuremath{\mathbf{adom}}}
\newcommand{\idom}{\ensuremath{\mathbf{idom}}}
\newcommand{\dc}{\ensuremath{\delta}}
\newcommand{\sel}{\ensuremath{\sigma}}
\newcommand{\vartrue}{\ensuremath{\texttt{T}}}
\newcommand{\varfalse}{\ensuremath{\texttt{F}}}
\newcommand{\pvar}{\ensuremath{\operatorname{\sf var}_\text{p}}}
\newcommand{\cvar}{\ensuremath{\operatorname{\sf var}_\text{v}}}

\newcommand{\idf}{\ensuremath{\operatorname{\sf idf}}}
\newcommand{\constr}{\ensuremath{\operatorname{\sf idf}}}
\newcommand{\prop}{\ensuremath{\operatorname{\sf prop}}}
\newcommand{\propth}{\ensuremath{\Pi}}
\newcommand{\pdc}{\ensuremath{\propth_\IC}}
\newcommand{\psel}{\ensuremath{\propth_\dec}}
\newcommand{\axioms}{\ensuremath{\propth_{\bot}}}

\newcommand{\IC}{\ensuremath{\Delta}}
\newcommand{\dec}{\ensuremath{\Sigma}}
\newcommand{\cdc}{\ensuremath{\operatorname{\sf cdc}}}
\newcommand{\cl}[1][\IC]{\ensuremath{{{#1}^*}}}
\newcommand{\cplex}[1]{\textsf{#1}}
\newcommand{\rules}[1]{\ensuremath{\mathcal{#1}}}
\newcommand{\cls}[1]{\textsf{#1}}

\newcommand{\iattr}[1]{\ensuremath{\|#1\|}}
\newcommand{\cond}{\ensuremath{\lambda}}
\newcommand{\pcond}{\ensuremath{P}}
\newcommand{\pidf}{\ensuremath{v}}
\newcommand{\phitilde}{\ensuremath{\widetilde{\varphi}}}

\begin{document}
%
\title{Lossless Selection Views\\under Conditional Domain Constraints%
  \thanks{This is an author version of the article published in \emph{IEEE
      Transactions on Knowledge and Data Engineering, Vol.\ 27, No.\ 2, pp.\
      504--517, IEEE, February 2015}. Digital Object Identifier:
    \href{http://dx.doi.org/10.1109/TKDE.2014.2334327}{%
      10.1109/TKDE.2014.2334327}}%
  \thanks{A preliminary version of this article appeared in \emph{Big Data:
      Proceedings of the 29th British National Conference on Databases, Vol.\
      7968 of LNCS, pp.\ 77--91, Springer, 2013}, under the title ``Lossless
    horizontal decomposition with domain constraints on interpreted
    attributes''.}}

%
%
%
%

\author{Ingo~Feinerer, Enrico~Franconi and Paolo~Guagliardo%
  \IEEEcompsocitemizethanks{%
    \IEEEcompsocthanksitem Ingo~Feinerer is with the Database and Artificial
    Intelligence Group, Institute of Information Systems, Vienna University of
    Technology, Austria. %
    E-mail: \texttt{ingo.feinerer@tuwien.ac.at} This work was supported by the
    Austrian Science Fund (FWF), project P25207-N23. %
    \IEEEcompsocthanksitem Enrico~Franconi and Paolo~Guagliardo are with the
    KRDB Research Centre, Free University of Bozen-Bolzano, Italy.
    \protect\\
    E-mail: \texttt{\{franconi,guagliardo\}@inf.unibz.it}}}

%
%

\markboth{Full version with complete proofs}{%
  Instead of this version, please cite
  \href{http://dx.doi.org/10.1109/TKDE.2014.2334327}{%
    \MakeLowercase{http://dx.doi.org/10.1109/}10.1109/TKDE.2014.2334327}}
%


\IEEEcompsoctitleabstractindextext{%
\begin{abstract}
  A set of views defined by selection queries splits a database relation into
  sub-relations, each containing a subset of the original rows. %
  This \emph{decomposition} into horizontal fragments is \emph{lossless} when
  the initial relation can be reconstructed from the fragments by union. %
  In this paper, we consider horizontal decomposition in a setting where some of
  the attributes in the database schema are \emph{interpreted} over a specific
  domain, on which a set of special predicates and functions is defined.

  We study losslessness in the presence of integrity constraints on the database
  schema. %
  We consider the class of \emph{conditional domain constraints} (CDCs), which
  restrict the values that the interpreted attributes may take whenever a
  certain condition holds on the non-interpreted ones, and investigate lossless
  horizontal decomposition under CDCs in isolation, as well as in combination
  with functional and unary inclusion dependencies.
\end{abstract}



\IEEEkeywords %
selection, views, losslessness, constraints, CDC, consistency, separability}

\maketitle

\IEEEdisplaynotcompsoctitleabstractindextext

%
\IEEEpeerreviewmaketitle

%
\ifCLASSOPTIONcompsoc
 \noindent\raisebox{2\baselineskip}[0pt][0pt]%
 {\parbox{\columnwidth}{\section{Introduction}\label{sec:introduction}%
 \global\everypar=\everypar}}%
 \vspace{-1\baselineskip}\vspace{-\parskip}\par
\else
 \section{Introduction}\label{sec:introduction}\par
\fi
%

%
%
%

\IEEEPARstart{T}{he} problem of updating a database through a set of views
consists in propagating updates issued on the views to the underlying base
relations over which the view relations are defined, so that the changes to the
database reflect \emph{exactly} those to the views. %
This is a classical problem in database research, known as the \emph{view update
  problem} (\cite{Bancilhon:1981:USR,Dayal:1982:CTU,Keller:1985:URD}), which in
recent years has received renewed and increasing attention
(\cite{Peng:2013:SEF,Caroprese:2012:VPI,Franconi:2012:TVU,Johnson:2008:CCR,%
  Bohannon:2006:RLL,Hegner:2004:OTU}).

View updates can be consistently propagated in an unambiguous way under the
condition that the mapping between database and view relations is
\emph{lossless}, which means that not only do the view relations depend on the
database relations, but also the converse is true. %
However, just knowing that such an ``inverse'' dependency exists is not yet
sufficient to effectively propagate the changes from the views to the
database. %
What is essential to know is \emph{how}, in some constructive way, the database
relations depend on the view relations. %
This amounts to being able to define each database relation in terms of the
views by means of a query, in much the same way the latter are defined from the
former \cite{Franconi:2013:EUC}. %
In such a context, database decompositions \cite{Abiteboul:1995:FD} play an
important role, because their losslessness is associated with the existence of
an explicit \emph{reconstruction operator} that, as the name suggests,
prescribes how a database relation can be rebuilt from the pieces, called
\emph{fragments}, into which it has been decomposed.

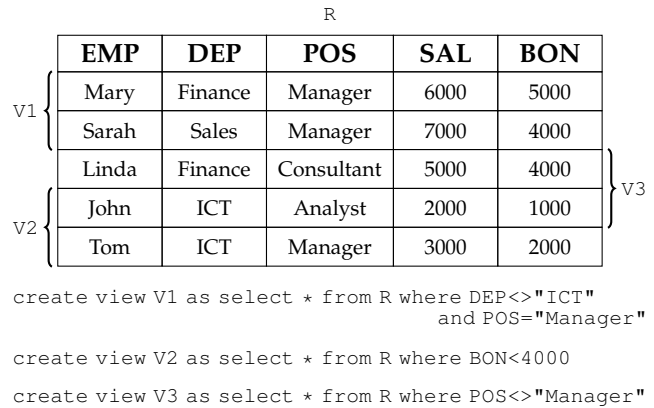
\begin{figure}
  \centering
  \begin{tikzpicture}[viewdef/.style={%
    font=\ttfamily,align=left,text width=.9531\linewidth}]%
  \fontsize{8pt}{8pt}
  \matrix [table=1.2cm, label={above:\texttt{R}}, %
  row 1/.style={font=\fontsize{10pt}{10pt}\bfseries\selectfont,%
    nodes={text depth=1pt}},
  column 3/.style={nodes={text width=1.5cm}}] (R) {
    EMP & DEP & POS & SAL & BON\\
    Mary & Finance & Manager & 6000 & 5000\\
    Sarah & Sales & Manager & 7000 & 4000\\
    Linda & Finance & Consultant & 5000 & 4000\\
    John & ICT & Analyst & 2000 & 1000\\
    Tom & ICT & Manager & 3000 & 2000\\
  };
  \draw [decorate,decoration={brace,mirror,raise=2pt},thick] %
  (R-2-1.north west) -- (R-3-1.south west) %
  node [midway, left, xshift=-4pt] {\texttt{V1}};
  \draw [decorate,decoration={brace,mirror,raise=2pt},thick] %
  (R-5-1.north west) -- (R-6-1.south west) %
  node [midway, left, xshift=-4pt] {\texttt{V2}};
  \draw [decorate,decoration={brace,raise=2pt},thick] %
  (R-4-5.north east) -- (R-5-5.south east) %
  node [midway, right, xshift=+4pt] {\texttt{V3}};

  \node [viewdef, below=1mm of R] (V1) {%
    create view V1 as select * from R where DEP<>"ICT"\\
    \hfill and POS="Manager"};

  \node [viewdef, below=1mm of V1] (V2) {%
    create view V2 as select * from R where BON<4000};

  \node [viewdef, below=1mm of V2] (V3) {%
    create view V3 as select * from R where POS<>"Manager"};
\end{tikzpicture}%
  \caption{Selection views over a company database.}
  \label{fig:hdec-example}
\end{figure} 

Lossless database decomposition is particularly relevant in distributed
settings, where fragments are scattered over a number of sites (typically within
a network), for the reason that it increases the throughput of the system by
allowing the concurrent execution of transactions as well as the parallel
execution of a single query as a set of subqueries operating on fragments
\cite{OzsuV:2011:pdds3}.

\emph{Horizontal decomposition} is the process of splitting a giv\-en relation
into sub-relations on the same attributes and of the same arity, each containing
a subset of the rows of the original relation.
For example, consider the relation \texttt{R} shown in
Figure~\ref{fig:hdec-example}, recording data about the employees of a company:
their name (\textbf{EMP}), the department (\textbf{DEP}) and the position
(\textbf{POS}) in which they are employed, and their income (e.g., euros per
month) consisting of a fixed salary (\textbf{SAL}) plus a variable bonus
(\textbf{BON}). %
In Figure~\ref{fig:hdec-example}, the relation \texttt{R} is decomposed into
three fragments: \texttt{V1} selects the rows of \texttt{R} with employees
working as managers in departments other than ICT, \texttt{V2} selects the rows
of \texttt{R} with employees who get strictly less than 4000 as bonus, and
\texttt{V3} selects the rows of \texttt{R} with employees who do not work as
managers. %
This kind of decomposition is lossless when the original relation can be
reconstructed from the fragments by union; in other words, the reconstruction
operator for horizontal decomposition is the union. %
In the example of Figure~\ref{fig:hdec-example}, the set of views $\{
\texttt{V1}, \texttt{V2}, \texttt{V3} \}$ constitutes a lossless horizontal
decomposition of \texttt{R}, as the union of \texttt{V1}, \texttt{V2} and
\texttt{V3} contains all (and only) the rows of \texttt{R}. %
Each proper subset of $\{ \texttt{V1}, \texttt{V2}, \texttt{V3} \}$ constitutes
a lossy decomposition of \texttt{R}, because each view selects at least one row
that is not selected by any of the others; e.g., the union of \texttt{V1} and
\texttt{V2} does not contain the third row of \texttt{R}, $\langle \text{Linda},
\text{Finance}, \text{Consultant}, 5000, 4000 \rangle$, which is selected only
by \texttt{V3}.

Observe that the horizontal decomposition specified by the definitions of views
\texttt{V1}, \texttt{V2} and \texttt{V3} in Figure~\ref{fig:hdec-example} is
lossless for the given relation \texttt{R}, but this is not the case for every
relation (over the same attributes). %
For instance, the tuple $\langle \text{Sam}, \text{ICT}, \text{Manager}, 6000,
5000 \rangle$ is not selected by any of these views; indeed, every relation
containing a row for an employee who works as a manager in the ICT department
and receives a bonus greater than $4000$ would not be losslessly decomposed by
\texttt{V1}, \texttt{V2} and \texttt{V3}. %
In the presence of integrity constraints, however, things may be different,
because some tuples, such as the one above, might not be allowed in the input
relation.

The study of horizontal decomposition (\cite{DeBra:1987:hdr,%
  DeBra:1986:icdt,CeriNP:1982:sigmod,MaierU:1983:sigmod}) has mostly focused on
settings where data values can only be compared for equality.
However, most real-world applications make use of data values coming from
domains with a richer structure (e.g., ordering) on which a variety of other
restrictions besides equality can be expressed (e.g., that of being within a
range or above a threshold). %
Examples are the attributes \textbf{SAL} and \textbf{BON} of the relation in
Figure~\ref{fig:hdec-example}, the dimensions, weights and prices in the
database of a shipping company, or the various amounts (credits, debits,
interest and exchange rates, etc.) recorded in a banking application.
It is therefore of practical interest to consider a scenario where some of the
attributes in the database schema are \emph{interpreted} over a specific domain,
such as the reals or the integers, on which a set of predicates (e.g.,
smaller/greater than) and functions (e.g., addition and subtraction) are
defined, according to a first-order language $\lang{C}$.

In the present article, we consider horizontal decomposition in a setting with
interpreted attributes, in which fragments are defined by selection queries
consisting of a condition on the non-interpreted attributes, expressed by a
Boolean combination of equalities, and a condition on the interpreted
attributes, expressed by a formula in $\lang{C}$. %
In particular, we study the \emph{losslessness} (w.r.t.\ every input relation)
of horizontal decompositions specified in this way, in the presence of integrity
constraints on the database schema.
We work under the pure universal relation assumption (URA)
\cite{Abiteboul:1995:FD}, that is, we restrict ourselves to a database schema
consisting of only one relation symbol, as customary in the study of database
decomposition.

\subsection*{Contribution and Outline}

In Section~\ref{sec:prelim}, we introduce a class of integrity constraints
called \emph{conditional domain constraints} (CDCs). %
By means of a formula in $\lang{C}$, a CDC restricts the values that the
interpreted attributes can take whenever a certain condition is satisfied by the
non-interpreted ones. %
Depending on the expressive power of $\lang{C}$, CDCs can capture constraints
that naturally arise in practise; for example, in the scenario of
Figure~\ref{fig:hdec-example}, it may be required that employees in the ICT
department have a total income (i.e., salary plus bonus) of at most 5000, that
employees working as managers get a bonus of at least 2000, and that employees
never receive a bonus greater than their salary. %
These constraints can be expressed as:
\begin{subequations}
  \begin{alignat}{5}
    \label{eq:inf-cdc1}
    \text{DEP} &= \text{``ICT''} & &\implies {} & \text{SAL} + \text{BON}& &
    &\le {} & 5000& \enspace ;\\
    \label{eq:inf-cdc2}
    \text{POS} &= \text{``Manager''} & &\implies & \text{BON}& & &\ge &
    2000& \enspace ;\\
    \label{eq:inf-cdc3}
    && && \text{SAL} - \text{BON}& & &\ge & 0& \enspace .
  \end{alignat}
\end{subequations}
As we shall see, the views of Figure~\ref{fig:hdec-example} losslessly decompose
every relation satisfying the above CDCs.

In our investigation, we do not commit to any specific language $\lang{C}$ and
we simply assume that $\lang{C}$ is closed under negation.

In Section~\ref{sec:consistency}, we characterise consistent sets of CDCs in
terms of satisfiability in $\lang{C}$. %
Whenever the satisfiability of sets of formulae in $\lang{C}$ is decidable, our
characterisation directly gives a decision procedure for checking whether a set
of CDCs is consistent. %
This is the case, e.g., for the so-called \emph{Unit Two Variable Per
  Inequality} fragment of linear arithmetic over the integers, whose formulae
(referred to as UTVPIs) consist of at most two variables and variables have unit
coefficients, as well as for Boolean combinations of such formulae. %
We prove that deciding consistency is \cplex{NP}-complete for both of these
languages.

In Section~\ref{sec:losslessness}, we characterise lossless horizontal
decomposition under CDCs in terms of unsatisfiability in $\lang{C}$. %
Whenever the satisfiability of sets of formulae in $\lang{C}$ is decidable, this
characterisation gives a decision procedure for checking whether a horizontal
decomposition is lossless under CDCs. %
We show that this problem is \cplex{co-NP}-complete when $\lang{C}$ is the
language of either UTVPIs or Boolean combinations of UTVPIs.

In Section~\ref{sec:separability}, we study lossless horizontal decomposition
under CDCs in combination with traditional integrity constraints. %
We show that functional dependencies (FDs) do not interact with CDCs and can
thus be allowed without any restriction, whereas this is not the case for unary
inclusion dependencies (UINDs). %
We provide a \emph{domain propagation rule} to derive a set of CDCs that fully
captures the interaction between a given set of UINDs and opportunely restricted
CDCs w.r.t.\ lossless horizontal decomposition, which makes possible to employ
the general technique for deciding losslessness also in the presence of UINDs. %
In addition, we consider restricted combinations of CDCs with both FDs and
UINDs.

We conclude in Section~\ref{sec:conclusion} with a discussion of the results,
relevant related work and future research directions.



\section{Preliminaries}
\label{sec:prelim}

We start by introducing the necessary notation and notions that will be used
throughout the article. %
We assume some familiarity with formal logic and its application to database
theory.

\smallskip

\noindent\textbf{Basics.}
An \emph{$n$-tuple} is an ordered list of $n$ elements, where $n$ is a positive
integer. %
We denote tuples by overlined lowercase letters (e.g., $\overline{t}$) and we
write them as comma-separated sequences enclosed in parentheses; the $k$-th
element of a tuple $\overline{t}$ is denoted by $\overline{t}[k]$. %
For example, if $\overline{t}$ is the 4-tuple $(a,b,c,a)$, then $\overline{t}[3]
= c$. %
An \emph{$n$-ary relation} on a set $A$, where $n$ is called the \emph{arity} of
the relation, is a set of $n$-tuples of elements of $A$.

A \emph{schema} is a finite set $\sign{S}$ of relation symbols, also called a
\emph{relational signature}. %
Each relation symbol $S$ has a positive arity $\arity{S}$ indicating the total
number of \emph{positions} in $S$, which are partitioned into \emph{interpreted}
and \emph{non-interpreted} ones. %
Relation symbols of arity $n$ are called \emph{$n$-ary}; we indicate that
$\arity{S} = n$ by writing $S/n$. %

Let $\dom$ be a possibly infinite set of arbitrary values, and let $\idom$ be a
set of values from a specific domain (e.g., the integers $\mathbb{Z}$) on which
a set of predicates (e.g., $\le$) and functions (e.g., $+$) are defined,
according to a first-order language $\lang{C}$ closed under negation. %
An \emph{instance} over a schema $\sign{S}$ associates each $S \in \sign{S}$
with a relation $S^I$ of appropriate arity on $\dom \cup \idom$, called the
\emph{extension} of $S$ under $I$, such that the values for the interpreted and
non-interpreted positions of $S$ are taken from $\idom$ and $\dom$,
respectively. %
The set of elements of $\dom \cup \idom$ occurring in an instance $I$ is called
the \emph{active domain} of $I$, denoted by $\adom(I)$. %
An instance is finite if its active domain is, and all instances in this article
are assumed to be finite. %
A \emph{fact} is given by the association, denoted by $R(\overline{t})$, between
a relation symbol $R$ and a tuple $\overline{t}$ of values of appropriate arity;
an instance can be represented as a set of facts.

\smallskip

\noindent\textbf{Constraints.}
A \emph{language} over a relational signature $\sign{S}$ is a set of first-order
logic (FOL) formulae over $\sign{S}$ with constants $\dom \cup \idom$ under the
standard name assumption (i.e., the interpretation of each constant is the
constant's name itself).
A formula in some language $\lang{L}$ is called an $\lang{L}$-formula. %
The sets of constants and relation symbols that occur in a formula $\varphi$ are
denoted by $\const(\varphi)$ and $\sig(\varphi)$, respectively; we extend
$\const(\cdot)$ and $\sig(\cdot)$ to sets of formulae in the natural way.

A \emph{constraint} is a closed formula (that is, without free variables) in
some language. %
For a set $\Gamma$ of constraints, we say that an instance $I$ over
$\sig(\Gamma)$ is a \emph{model} of (or \emph{satisfies}) $\Gamma$, and write $I
\models \Gamma$, to indicate that the relational structure $\langle \adom(I)$
$\cup \const(\Gamma)$, $I \rangle$ makes every formula in $\Gamma$ true under
the standard FOL semantics. %
We write $I \models \varphi$ as short for $I \models \{\varphi\}$, and say that
$I$ satisfies $\varphi$. %
A set of constraints $\Gamma$ \emph{entails} (or \emph{logically implies}) a
constraint $\varphi$, written $\Gamma \models \varphi$, if every finite model of
$\Gamma$ also satisfies $\varphi$. %
All sets of constraints in this article are finite.


\smallskip

\noindent\textbf{Propositional Theories.}
A \emph{propositional variable} is a variable whose value can be either
$\vartrue$ (true) or $\varfalse$ (false). %
A \emph{propositional formula} is a Boolean combination of \emph{propositional
  variables}, including the two special propositional variables $\top$ and
$\bot$, whose values are always $\vartrue$ and $\varfalse$, respectively. %
A \emph{propositional theory} is a set of propositional formulae. %
We denote the set of propositional variables occurring in a propositional
formula $P$ by $\var(P)$ and we extend $\var(\cdot)$ to propositional theories
in the natural way. %
A \emph{valuation} of a set of propositional variables (also called a
\emph{truth-value assignment}) assigns a \emph{truth-value} (i.e., either
$\vartrue$ or $\varfalse$) to each propositional variable in the set. %
The truth-value $\alpha(P)$ of a propositional formula $P$ under a valuation
$\alpha$ of its propositional variables is determined by the standard semantics
of the Boolean connectives. %
We say that $\alpha$ \emph{satisfies} (or \emph{makes true}) $P$, and write
$\alpha \models P$, if $\alpha(P) = \vartrue$. %
Given a propositional theory $\Pi$, a valuation of $\var(\Pi)$ \emph{satisfies}
$\Pi$, written $\alpha \models \Pi$, if $\alpha$ satisfies every propositional
formula in $\Pi$.

\subsection*{Horizontal Decomposition}

We consider a \emph{source schema} $\R$, consisting of a single relation symbol
$R$, and a \emph{decomposed schema} $\V$, disjoint with $\R$, of \emph{view}
symbols with the same arity as $R$. %
We formally define horizontal decomposition as follows.

\begin{definition}
  Let $\R = \{ R \}$ and $\V = \{ V_1, \dotsc, V_n \}$. %
  Let $\IC$ be a set of constraints over $\R$ and let $\dec$ be a set of exact
  view definitions, one for each $V_i \in \V$, of the form $\forall \overline{x}
  \mathinner.  V_i(\overline{x}) \leftrightarrow \varphi(\overline{x})$, where
  $\varphi$ is a safe-range\footnotemark\ formula over $\R$. %
  Then, $\dec$ is a \emph{horizontal decomposition of $\R$ into $\V$ under
    $\IC$} if ${\IC \cup \dec} \models \forall \overline{x} \mathinner.
  V_i(\overline{x}) \rightarrow R(\overline{x})$ for every $V_i \in \V$. %
  We say that $\dec$ is \emph{lossless} if ${\IC \cup \dec} \models {\forall
    \overline{x} \mathinner. R(\overline{x}) \leftrightarrow {V_1(\overline{x})
      \lor \dotsb \lor V_n(\overline{x})}}$.
\end{definition} 

\footnotetext{For details on the syntactic notion of range restriction,
  corresponding to the semantic notion of domain independence, refer to
  \cite{Abiteboul:1995:FD}.}

For the sake of simplicity, w.l.o.g.\ we assume that the first $\iattr{R}$
positions of $R$ and of every $V \in \V$ are non-interpreted, while the
remaining $\arity{R} - \iattr{R}$ positions are interpreted. %
Under this assumption, instances over $\R \cup \V$ associate each relation
symbol with a subset of the Cartesian product $\dom^k \times \idom^{n-k}$, where
$n = \arity{R}$ and $k = \iattr{R}$.
Unless otherwise specified, when we speak of a tuple $\overline{t}$ we
implicitly assume that $\overline{t}$ is of arity $n$ and that the first $k$
values of $\overline{t}$ are from $\dom$ while the rest are from $\idom$.
W.l.o.g.\ we also assume that a variable associated with the $i$-th position of
$R$ is named $x_i$ if $i \le k$, and $y_{i-k}$ otherwise. %
By default, $\overline{x}$ and $\overline{y}$ denote the tuples $(x_1, \dots,
x_k)$ and $(y_1, \dots, y_{n-k})$, respectively.

Since every $\lang{C}$-formula is over variables associated with interpreted
positions, we write $\phi(\overline{y})$ to indicate that $\phi$ is a
$\lang{C}$-formula whose free variables are among the variables in
$\overline{y}$. %
For a tuple $\overline{t}_y$ of $n-k$ values from $\idom$, we denote by
$\phi(\overline{t}_y)$ the result of replacing every occurrence in $\phi$ of the
free variable $y_i$ with the value $\overline{t}_y[i]$. %
We say that $\overline{t}_y$ is a \emph{solution} to $\phi$ if
$\phi(\overline{t}_y)$ is true under the semantics of $\lang{C}$. %
In such a case, we also say that the assignment $\beta$ associating each $y_i$
with $\overline{t}_y[i]$ \emph{satisfies} $\phi$, and we write $\beta \models
\phi$.\footnote{Sometimes, by abuse of terminology, we say that an assignment
  $\beta$ is a solution to a $\lang{C}$-formula $\phi$, with the obvious
  meaning.}

\smallskip

\noindent\textbf{Source Constraints.}
The class of integrity constraints we consider on the source schema $\R$ is that
of \emph{conditional domain constraints} (CDCs), which restrict the admissible
values at interpreted positions by means of formulae in $\lang{C}$, when a
certain condition holds on the non-interpreted ones. %
Formally, a CDC is a formula of the form
\begin{equation}
  \label{eq:cdc}
  \forall \overline{x}, \overline{y} \mathinner. %
    {\bigl( R(\overline{x},\overline{y}) \land \cond(\overline{x}) \bigr) %
    \rightarrow \dc(\overline{y})} %
  \enspace ,
\end{equation} 
where $\cond(\overline{x})$ is a Boolean combination of equalities $x=a$, with
$x $ from $\overline{x}$ and $a$ from $\dom$, and $\dc(\overline{y}) \in
\lang{C}$. %
We use $x \neq a$ as short for $\lnot (x=a)$ and, for ease of notation, we write
\eqref{eq:cdc} simply as $\cond(\overline{x}) \rightarrow \dc(\overline{y})$. %
Here, we make use of a more general variant of the CDCs introduced in
\cite{FeinererFG:2013:bncod}, where the condition $\cond(\overline{x})$ was
limited to a conjunction of possibly negated equalities. %
In general, \eqref{eq:cdc} is more expressive than a CDC of the form used in
\cite{FeinererFG:2013:bncod}, as in the latter negation is allowed only
atomically, that is, in front of equalities, and so disjunction cannot be
expressed. %
However, there is no difference in expressivity between the two variants when
considering sets of CDCs, because the antecedent of \eqref{eq:cdc} can always be
rewritten in disjunctive normal form (DNF) and the CDC split into a set of CDCs
having the same consequent and each disjunct as antecedent.
E.g., the CDC ${x_1 = a} \land ({x_2 \neq b} \lor {x_3 = c}) \rightarrow
\dc(\overline{y})$ is equivalent to the following set of CDCs:
\begin{equation*}
  \bigl\{ %
  {x_1 = a} \land {x_2 \neq b} \rightarrow \dc(\overline{y}),\; %
  {x_1 = a} \land {x_3 = c} \rightarrow \dc(\overline{y}) %
  \bigr\} \enspace .
\end{equation*}

Standard domain constraints on non-interpreted attributes, of the form
$R(\overline{x},\overline{y}) \rightarrow {x_i = a_1} \lor \dotsb \lor {x_i =
  a_n}$ with $a_1, \dotsc, a_n \in \dom$, can be expressed by the two following
CDCs,\footnote{Repetition of variables in the antecedent of a CDC is allowed.}
for some $\dc(\overline{y}) \in \lang{C}$:\footnote{Recall that $\lang{C}$ is
  closed under negation, hence $\lnot \dc(\overline{y}) \in \lang{C}$.}
\begin{align*}
  {x_i \neq a_1} \land \dotsb \land {x_i \neq a_n} %
  &\rightarrow \phantom{\lnot}\dc(\overline{y}) \enspace ;\\
  {x_i \neq a_1} \land \dotsb \land {x_i \neq a_n} %
  &\rightarrow {\lnot \dc(\overline{y})} \enspace .
\end{align*} 

\smallskip

\noindent\textbf{View Definitions.}
The view symbols in $\V$ are defined by selection queries with conditions on
both interpreted and non-interpreted positions. %
Formally, each $V \in \V$ is defined by a formula of the form
\begin{equation}
  \label{eq:sel}
  \forall \overline{x},\overline{y} \mathinner. V(\overline{x},\overline{y}) %
  \leftrightarrow \bigl( R(\overline{x},\overline{y}) \land \cond(\overline{x})
  \land \sel(\overline{y}) \bigr) \enspace ,
\end{equation} 
where $\cond(\overline{x})$ is as in \eqref{eq:cdc} and $\sel(\overline{y}) \in
\lang{C}$. %
In the following, we write \eqref{eq:sel} simply as $V \colon
{\cond(\overline{x}) \land \sel(\overline{y})}$. %
View definitions of this form clearly generalise those in
\cite{FeinererFG:2013:bncod}, where $\cond(\overline{x})$ is limited to a
conjunction of possibly negated equalities and disjunction cannot be
expressed. %
While this has an impact on end-users, who can define more expressive views,
there is no difference between the two formalisms w.r.t.\ losslessness, in that
any view symbol $V$ defined by \eqref{eq:sel} can be split into a set of views,
defined by formulae of the form used in \cite{FeinererFG:2013:bncod}, that
together select exactly the same tuples as $V$; given $\cond(\overline{x})$ in
DNF, each of these view definitions has the same selection condition as $V$ on
the interpreted attributes and a disjunct of $\cond(\overline{x})$ as selection
condition on the non-interpreted ones.
For example, the view $V \colon {x_1 = a} \land ({x_2 \neq b} \lor {x_3 = c})
\land \sel(\overline{y})$ selects the same tuples selected by the following set
of views:
\begin{equation*}
  \bigl\{ %
  V_1 \colon {x_1 = a} \land {x_2 \neq b} \land \sel(\overline{y}),\, %
  V_2 \colon {x_1 = a} \land {x_3 = c} \land \sel(\overline{y}) %
  \bigr\} \; .
\end{equation*}

The technique we will present in Section~\ref{sec:losslessness} for checking
whether a set of selections of the form \eqref{eq:sel} is lossless can also be
applied when (some of) the selections have the form $V \colon
\cond(\overline{x}) \lor \sel(\overline{y})$ by considering, in place of each
such selection, the two selections $V' \colon \cond(\overline{x})$ and $V''
\colon \sel(\overline{y})$.

\smallskip

\noindent\textbf{Running Example.}
To clarify the notation and illustrate the concepts introduced so far, we now
give an example that will be used also in the rest of the article. %
It is based on the source schema of Figure~\ref{fig:hdec-example}, the CDCs
\eqref{eq:inf-cdc1}--\eqref{eq:inf-cdc3} informally described in
Section~\ref{sec:introduction} and the views previously specified in
Figure~\ref{fig:hdec-example} by means of SQL statements.

\begin{example}
  \label{exa:running}
  Let $R$ be a relation symbol of arity $5$, whose positions are associated with
  attributes \textbf{EMP}, \textbf{DEP}, \textbf{POS}, \textbf{SAL},
  \textbf{BON}, in this order, with the last two interpreted over the
  integers. %
  Differently from the example of Figure~\ref{fig:hdec-example}, for simplicity
  we assume that salaries and bonuses are given in thousands of euros/month. %
  Let $a = \text{``ICT''}$ and $b = \text{``Man\-ager''}$, and consider the
  following set $\IC$ of CDCs:
  \begin{subequations}
    \begin{alignat}{2}
      x_2 = a \rightarrow {}&& y_1 + y_2 \le 5 \enspace ;\\
      x_3 = b \rightarrow {}&&       y_2 \ge 2 \enspace ;\\
      \top \rightarrow {}&& y_1 - y_2 \ge 0 \enspace .
    \end{alignat}
  \end{subequations} 

  Let $\V = \{\, V_1, V_2, V_3 \,\}$, and consider the horizontal
  de\-composition $\dec$ given by
  \begin{flalign*}
    && V_1 \colon {{x_2 \neq a} \land {x_3 = b}} \enspace ; &&
    V_2 \colon {y_2 < 4} \enspace ; &&
    V_3 \colon {x_3 \neq b} \enspace .&&
  \end{flalign*} 
\end{example} 

\smallskip

\noindent\textbf{Specific Languages.}
The techniques we will present for deciding whether a set of CDCs is consistent
(Section~\ref{sec:consistency}) and whether a horizontal decomposition is
lossless under CDCs (Section~\ref{sec:losslessness}) give actual algorithms when
satisfiability in \lang{C} is decidable; in the case of losslessness, \lang{C}
is additionally required to be closed under negation. %
Thus, even though our investigation is in general independent of the choice of
\lang{C}, from a practical point of view it makes sense to consider concrete
languages that enjoy both of the above properties. %
Two prominent such languages are \emph{Unit Two-Variable Per Inequality}
formulae (UTVPIs) and Boolean combinations thereof. %
UTVPIs, a.k.a.\ \emph{Generalised $2$SAT} (G$2$SAT) formulae
\cite{Barrett:2009:SMT}, are a fragment of linear arithmetic over the
integers. %
Formally, a UTVPI formula has the form $ax + by \le d$, where $x$ and $y$ are
integer variables, $a,b \in \{ -1, 0, 1 \}$ and $d \in \mathbb{Z}$. %
The following equivalences hold:
\begin{subequations}
  \begin{alignat}{6}
    ax + by \ge d \iff {}& & (-ax&) & {} + {}& & (-by&) & &\le (-d) %
    &&\enspace ; \\
    ax + by < d \iff {}& & ax& & {} + {}& & by& & &\le (d-1) %
    &&\enspace ; \\
    ax + by > d \iff {}& & ax& & {} + {}& & by& & &\ge (d+1) %
    &&\enspace .
  \end{alignat}
\end{subequations} 

Thus, UTVPIs can express comparisons between two variables and between a
variable and an integer, as well as compare the sum or difference of two
variables with an integer. %
As integers allow to represent also real numbers with fixed precision, UTVPIs
may be sufficient for most applications. %
The CDCs and the view definitions of Example~\ref{exa:running} can be expressed
when \lang{C} is the language of UTVPIs.

Observe that the equality $x = y$, where $y$ is a variable or an integer, is not
expressible within a single UTVPI; instead, a set consisting of two UTVPIs,
namely $x \le y$ and $y \ge x$, is required. %
Therefore, in the consequent of a CDC, equality between variables or between a
variable and an integer is expressed as follows:
\begin{align*}
  \cond(\overline{x}) \rightarrow {y = z} \iff %
  \phantom{\bigl\{}\cond(\overline{x}) &\rightarrow {y \le z} \land {y \ge z}
  \\
  \iff \bigl\{ \cond(\overline{x}) &\rightarrow {y \le z} \,,\; %
  \cond(\overline{x}) \rightarrow {y \ge z} \bigr\} \enspace ,
\end{align*} 
with $z$ either a variable or an integer. %
Equality between the sum or difference of two variables and an integer is
expressed in a similar way.

Whether a set of UTVPIs is satisfiable can be checked in polynomial time
(\cite{SchuttS:2010:informs,LahiriM:2005:frocos,JaffarMSY:1994:ppcp}). %
We refer to a Boolean combination of UTVPIs as BUTVPI; deciding the
satisfiability of a set of BUTVPIs is \cplex{NP}-complete
\cite{SeshiaSB:2007:jsat}.



\section{Consistent Sets of CDCs}
\label{sec:consistency}

Before turning our attention to horizontal decomposition, we first deal with the
relevant problem of determining whether a set of CDCs is \emph{consistent}, that
is, whether it has a non-empty model.\footnotemark\ %
It is important to make sure that the integrity constraints over the source
schema are consistent, as every horizontal decomposition is meaninglessly
lossless when there are in fact no legal relations to decompose.

\footnotetext{Since CDCs are universally-quantified closed implicational
  formulae, any set thereof is always trivially satisfied by the empty
  instance.}

In this section, we will characterise the consistency of a set of CDCs in terms
of satisfiability in $\lang{C}$, where $\lang{C}$ is \emph{not} required to be
closed under negation. %
The \emph{consistency problem} for CDCs is the decision problem that takes as
input a set $\IC$ of CDCs and answers the question: ``Is $\IC$ consistent?'' %
We will show that when $\lang{C}$ is the language of either UTVPIs or BUTVPIs
this problem is \cplex{NP}-complete. %
The technique employed here provides the basis for the approach we follow in
Section~\ref{sec:losslessness} in the study of lossless horizontal
decomposition.

Observe that, given their form, CDCs affect only one tuple at a time, and so
whether an instance satisfies a set of CDCs depends on each tuple of the
instance in isolation from the others. %
Indeed, a set of CDCs is consistent precisely if it is satisfiable on an
instance consisting of only one tuple, therefore we can restrict our attention
to single tuples. %
Moreover, we are not really interested in the actual values of a tuple at
non-interpreted positions; what we need to know is simply whether such values
satisfy the conditions in the antecedent of each CDC or not.
To this end, with each equality between a variable $x_i$ and a constant $a$ we
associate a propositional variable $p_i^a$, whose truth-value indicates whether
the value in the $i$-th position is $a$. %
To each valuation of such propositional variables corresponds the (possibly
infinite) set of tuples satisfying the equalities associated with the names of
the propositional variables. %
For example, a valuation assigning true to $p^a_1$ and false to $p^b_2$
identifies all the tuples in which the value of the first element is $a$ and the
value of the second is different from $b$.
A bit more care is needed with valuations of propositional variables that refer
to the same position (i.e., have the same subscript) but to different constants
(i.e., have different superscripts). %
For example, $p_1^a$ and $p_1^b$ (with $a \neq b$) should never be both
evaluated to true.

As we shall see, checking whether a set $\IC$ of CDCs is consistent amounts to
first building a propositional theory by replacing the equalities with the
corresponding propositional variables, and then looking for a valuation $\alpha$
such that:
\begin{itemize}
\item any two propositional variables referring to the same position but to
  different constants are not evaluated both to true; and
\item the set of $\lang{C}$-formulae that ``apply'' under $\alpha$ is
  satisfiable.
\end{itemize}

\begin{definition}
  \label{def:prop-IC}
  Let $\IC = \{ \phi_1, \dotsc, \phi_n \}$ be a set of CDCs over $R$. %
  For each $\phi_i \in \IC$, recalling it has the form \eqref{eq:cdc}, we
  construct
  \begin{equation}
    \prop(\phi_i) = \pcond \rightarrow \pidf_i \enspace ,
  \end{equation} 
  where $\pcond$ is a propositional formula (possibly $\top$) obtained from the
  condition $\cond(\overline{x})$ in the antecedent of $\phi$ by replacing each
  equality $x_i = a$ between a variable $x_i$ and $a \in \idom$ with the
  propositional variable $p_i^a$, and $\pidf_i$ is a fresh propositional
  variable associated with the $\lang{C}$-formula $\dc(\overline{y})$, denoted
  by $\constr(\pidf_i)$,\footnote{\constr\ stands for ``interpreted domain
    formula''.}  in the consequent of $\phi$. %
  We denote $\{ \prop(\phi) \mathrel{|} \phi \in \IC \}$ by $\pdc$ and we call
  it the \emph{propositional theory associated with $\IC$}.
\end{definition} 

We consider the set $\var(\pdc)$ of propositional variables occurring in $\pdc$
partitioned into $\pvar(\pdc) = \{ \var(\pcond) \mid (\pcond \rightarrow
\pidf_i) \in \pdc \}$ and $\cvar(\pdc) = \var(\pdc) \setminus \pvar(\pdc)$.

For a pair of distinct propositional variables $p^a_i$ and $p^b_i$ associated
with the same position $i$ but distinct $\dom$ constants $a$ and $b$, we
consider the propositional formula $p_i^a \land p_i^b \rightarrow \bot$, called
the \emph{axiom of unique value} for $p^a_i$ and $p^b_i$, intuitively stating
that two distinct constants are not allowed in the same position.
The \emph{axioms of unique value} for a set of propositional variables consist
of the axiom of unique value for each pair of distinct prop\-ositional variables
$p^a_i$ and $p^b_i$ in the set.
A tuple $\overline{t}$ is \emph{consistent with a valuation $\alpha$} if, for
every propositional variable $p_i^a$, it holds that $\overline{t}[i] = a$
precisely if $\alpha(p_i^a) = \vartrue$.
In general, given a valuation $\alpha$ of a set of propositional variables, by
construction there exists a tuple consistent with $\alpha$ if and only if
$\alpha$ satisfies the corresponding axioms of unique value for that set.

\renewcommand{\axioms}{\ensuremath{\Pi_{\text{aux}}}}

\begin{definition}
  \label{def:aux-theory}
  Let $\IC$ be a set of CDCs over $R$. %
  The \emph{auxiliary theory $\axioms$ for $\pdc$} consists of the axioms of
  unique value for $\pvar(\pdc)$.
\end{definition} 

\begin{example}
  \label{exa:utvpi-cdc-prop}
  The propositional theory associated with $\IC$ of Example~\ref{exa:running} is
  \begin{equation*}
    \pdc = \bigl\{\, %
    {p_2^a \rightarrow \pidf_1},\, %
    {p_3^b \rightarrow \pidf_2},\, %
    {\top \rightarrow \pidf_3}\, %
    \bigr\} \enspace ,
  \end{equation*}
  where $\pvar(\pdc) = \{\, p_2^a,\, p_3^b \,\}$ and $\cvar(\pdc) = \{ \pidf_1,
  \pidf_2, \pidf_3 \}$. %
  The auxiliary theory for $\pdc$ is $\axioms = \emptyset$. %
  The association between the propositional variables in $\cvar(\pdc)$ and the
  set of UTVPIs from the CDCs in $\IC$ is $\idf = \{\, \pidf_1 \mapsto y_1 + y_2
  \le 5,\, \pidf_2 \mapsto {y_2 \ge 2},\, \pidf_3 \mapsto {y_1 - y_2 \ge 0}
  \,\}$.
\end{example} 

Given a set $\IC$ of CDCs and a valuation $\alpha$ of $\pvar(\pdc)$, we say that
a CDC $\phi \in \IC$ is \emph{applicable} under $\alpha$ if $\alpha$ makes the
l.h.s.\ of $\prop(\phi)$ true. %
We can use $\alpha$ to ``filter'' $\pdc$ and construct a set consisting of the
consequent of each CDC in $\IC$ that is applicable under $\alpha$. %
This set contains the \lang{C}-formulae that must be necessarily satisfied by
the values at interpreted positions of every tuple consistent with $\alpha$,
that is, whose values at non-interpreted positions satisfy the antecedents of
the CDCs applicable under $\alpha$.

\begin{definition}
  Let $\IC$ consist of CDCs over $R$, and let $\alpha$ be a valuation of
  $\pvar(\pdc)$. %
  The \emph{$\alpha$-filtering of $\pdc$} is the set
  \begin{equation}
    \label{eq:alpha-cdc}
    \pdc^\alpha = \{\, \constr(\pidf) \mathrel{|} {} %
    (\pcond \rightarrow \pidf) \in \pdc,\, \alpha(\pcond) = \vartrue \,\}
    \enspace ,
  \end{equation} 
  consisting of $\lang{C}$-formulae associated with propositional variables that
  occur in some propositional formula of $\pdc$ whose l.h.s.\ holds true under
  $\alpha$.
\end{definition} 

The main result of this section characterises the consistency of a set of CDCs
in terms of satisfiability in $\lang{C}$. %
We remark again that the result holds in general for any language $\lang{C}$,
not necessarily closed under negation. %
This requirement will become essential only in the upcoming
Section~\ref{sec:losslessness} and Section~\ref{sec:separability}.

\begin{theorem}
  \label{thm:cdc-sat}
  Let $\IC$ be a set of CDCs over $R$, and let $\axioms$ be the auxiliary theory
  for $\pdc$. %
  Then, $\IC$ is consistent if and only if there exists a valuation $\alpha$ of
  $\pvar(\pdc)$ satisfying $\axioms$ and such that $\pdc^\alpha$ is satisfiable.
\end{theorem} 

Whenever the satisfiability of sets of $\lang{C}$-formulae is decidable,
Theorem~\ref{thm:cdc-sat} gives an algorithm to check whether a set of CDCs is
consistent, as we illustrate below in our running example, where $\lang{C}$ is
the language of UTVPIs.

\begin{example}
  \label{exa:utvpi-cdc-sat}
  With respect to $\pdc$ of Example~\ref{exa:utvpi-cdc-prop}, consider the
  valuation $\alpha = \{ \, p_2^a \mapsto \vartrue,\, p_3^b \mapsto \varfalse
  \,\}$, for which we have $\pdc^\alpha = \{\, {y_1 + y_2 \le 5},\, {y_1 - y_2
    \ge 0} \,\}$. %
  Obviously, $\alpha$ satisfies the (empty) auxiliary theory $\axioms$ for
  $\pdc$. %
  In addition, $\pdc^\alpha$ is satisfiable as, e.g., $\{\, y_1 \mapsto 3,\, y_2
  \mapsto 2 \,\}$ is a solution to every UTVPI in it.
\end{example} 

We will now give the proof of Theorem~\ref{thm:cdc-sat}, for which we first need
to prove a technical lemma. %
Let $n = \arity{R}$ and $k = \iattr{R}$; with each tuple $\overline{t}$ is
associated the assignment $\beta \colon \{ y_1, \dots, y_{n-k}\} \to \idom$,
which we refer to as the \emph{assignment induced by the interpreted positions
  of $\overline{t}$}, such that $\beta(y_{i-k}) = \overline{t}[i]$ for every $i
\in \{k+1, \dotsc, n\}$.
%
%
Intuitively, the following lemma shows that any tuple that is consistent with a
valuation $\alpha$ satisfies a set of CDCs precisely if the assignment
induced by its interpreted positions satisfies the $\alpha$-filtering.

\begin{lemma}
  \label{lem:key-consistency}
  Let $\IC$ be a set of CDCs over $R$, and let $\alpha$ be a valuation of
  $\pvar(\pdc)$. %
  Let $\overline{t}$ be consistent with $\alpha$, and let $\beta$ be the
  assignment induced by the interpreted positions of $\overline{t}$. %
  Then, $\{ R(\overline{t}) \} \models {\IC}$ if and only if $\beta$ satisfies
  $\pdc^\alpha$.
\end{lemma} 

\begin{pf}
  Let $n = \arity{R}$ and $k = \iattr{R}$.

  \begin{enumerate}[claim]
  \item Let $\phi \in \IC$ and $\prop(\phi) = \pcond \rightarrow
    \pidf$. %
    Then, $\alpha(\pcond) = \vartrue$ iff $\cond(\overline{x})$ is true under
    $\{ x_1 \mapsto \overline{t}[1], \dotsc, x_k \mapsto \overline{t}[k] \}$.%
    \label{cla:key-consist-1}
  \end{enumerate}

  \begin{description}[claimproof]
  \item[Proof.] %
    Since $\overline{t}$ is consistent with $\alpha$, for $i \in \{ 1, \dotsc, k
    \}$ we have that $\overline{t}[i] = a$ if and only if $\alpha(p^a_i) =
    \vartrue$.
  \end{description}

  \begin{enumerate}[claim, resume]
  \item For each $\prop(\phi) = \pcond \rightarrow \pidf$ with $\phi
    \in \IC$, it is the case that $I \not\models \phi$ if and only if
    $\alpha(\pcond) = \vartrue$ and $\beta \not\models \constr(\pidf)$.%
    \label{cla:key-consist-2}
  \end{enumerate}

  \begin{description}[claimproof]
  \item[Proof.] %
    As $\phi$ is a CDC, $I \not\models \phi$ if and only if the antecedent
    $\cond(\overline{x})$ of $\phi$ holds true under $\{ x_1 \mapsto
    \overline{t}[1], \dotsc, x_k \mapsto \overline{t}[k] \}$ and the consequent
    $\dc(\overline{y})$ of $\phi$ is not true under $\{ y_1 \mapsto
    \overline{t}[k+1], \dotsc, y_{n-k} \mapsto \overline{t}[n] \}$. %
    In turn, this is the case if and only if both $\alpha(\pcond) = \vartrue$
    (by \ref{cla:key-consist-1}) and $\beta$ does not satisfy $\constr(\pidf) =
    \dc(\overline{y})$ (by construction).
  \end{description} 

  We prove Lemma~\ref{lem:key-consistency} by showing that $I \not\models \IC$
  if and only if $\beta$ does not satisfy $\pdc^\alpha$.

  \begin{description}[iff]
  \item[``if''.] Assume $\beta \not\models \pdc^\alpha$, that is, there is some
    $\lang{C}$-formula $\psi \in \pdc^\alpha$ not satisfied by $\beta$. %
    By construction of $\pdc^\alpha$, $\psi$ is the consequent of a CDC $\phi
    \in \IC$ such that $\prop(\phi) = \pcond \rightarrow \pidf$, with $\psi =
    \idf(\pidf)$ and $\alpha(\pcond) = \vartrue$. %
    Thus, as $\beta$ does not satisfy $\psi$, by \ref{cla:key-consist-2} $I
    \not\models \phi$, and therefore $I \not\models \IC$.

  \item[``only if''.] Assume $I \not\models \IC$. %
    Then, there exists some $\phi \in \IC$ which is not satisfied by $I$. %
    Since $\prop(\phi) = \pcond \rightarrow \pidf$, by \ref{cla:key-consist-2}
    $\alpha(\pcond) = \vartrue$ and $\beta \not\models \constr(\pidf)$. %
    Hence, $\constr(\pidf) \in \pdc^\alpha$. %
    Therefore, $\beta \not\models \pdc^\alpha$.\hfill\IEEEQED
  \end{description}
\end{pf} 

\begin{thmpf}[of Theorem~\ref{thm:cdc-sat}]
  Let $n = \arity{R}$ and $k = \iattr{R}$.

  \begin{description}[iff]
  \item[``if''.] Let $\alpha$ and $\beta$ be such that $\alpha \models\axioms$
    and $\beta \models \pdc^\alpha$. %
    Then, as $\alpha \models \axioms$, it is never the case that two distinct
    propositional variables in $\pvar(\pdc)$ associated with the same position
    are both true under $\alpha$. %
    Thus, there exists a tuple $\overline{t}$ consistent with $\alpha$ and such
    that $\beta$ is the assignment induced by its interpreted positions. %
    Therefore, as $\beta \models \pdc^\alpha$, the instance $\{ R(\overline{t})
    \}$ is a model of $\IC$ by Lemma~\ref{lem:key-consistency}.

  \item[``only if''.]  Assume that $\IC$ is consistent, that is, it has a
    non-empty model. %
    In particular, as every formula in $\IC$ is in one tuple, there is a
    tuple $\overline{t}$ such that the instance $I = \{ R(\overline{t}) \}$ is a
    model of $\IC$. %
    Take $\alpha$ as follows: for every propositional variable $p \in
    \pvar(\pdc)$, $\alpha(p) = \vartrue$ if $p = p_i^a$ and $\overline{t}[i] =
    a$, otherwise $\alpha(p) = \varfalse$. %
    By construction, $\alpha \models \axioms$ and $\overline{t}$ is consistent
    with $\alpha$. %
    Therefore, as $I \models \IC$, the assignment $\beta$ induced by the
    interpreted positions of $\overline{t}$ satisfies $\pdc^\alpha$ by
    Lemma~\ref{lem:key-consistency}.\hfill\IEEEQED
  \end{description}
\end{thmpf} 

The satisfiability problem for $\lang{C}$ takes as input a set $\Gamma$ of
$\lang{C}$-formulae and answers the question: ``Is $\Gamma$ satisfiable?''

\begin{lemma}
  \label{lem:sat-to-consistency}
  The satisfiability problem for $\lang{C}$ linearly reduces to the consistency
  problem for CDCs.
\end{lemma} 

\begin{pf}
  Let $\Gamma = \{ \phi_1, \dotsc, \phi_n \}$ be a set $\lang{C}$-formulae. %
  Then, take $\IC = \{ \top \rightarrow \phi_i \mid \phi_i \in \Gamma \}$, let
  $\pdc = \{ \top \rightarrow v_i \mid i = 1, \dotsc, n \}$ and $\idf = \{ v_i
  \mapsto \phi_i \mid i = 1, \dotsc, n \}$. %
  The auxiliary theory for $\pdc$ is $\axioms = \emptyset$. %
  As $\pvar(\pdc) = \emptyset$, the only valuation of $\pvar(\pdc)$ is $\alpha =
  \emptyset$, which satisfies $\axioms$ and for which $\pdc^\alpha = \{\,
  \idf(v_i) \mid v_i \in \cvar(\pdc) \,\} = \Gamma$. %
  Thus, by Theorem~\ref{thm:cdc-sat}, the set $\IC$ of CDCs is consistent iff
  $\Gamma$ is satisfiable. %
  The reduction is linear in the size of $\Gamma$.\hfill\IEEEQED
\end{pf}

With regard to the consistency problem for CDCs whose consequents are either
UTVPIs or BUTVPIs, we have the following complexity results.

\begin{theorem}
  \label{thm:consistency-probl}
  When $\lang{C}$ is the language of either BUTVPIs or UTVPIs, the consistency
  problem for CDCs is \cplex{NP}-complete.
\end{theorem} 

\begin{pf}
  Constructing the propositional theories $\pdc$ and $\axioms$ requires linear
  time, checking that a valuation $\alpha$ of $\pvar(\pdc)$ satisfies $\axioms$
  takes polynomial time, and checking that an assignment from the variables in
  $\overline{y}$ to integers satisfies $\pdc^\alpha$ (whose construction takes
  linear time) can be done in polynomial time, whether $\pdc^\alpha$ consists of
  either UTVPIs or BUTVPIs. %
  Hence, in light of Theorem~\ref{thm:cdc-sat}, we can verify a given solution
  to the consistency problem, when $\lang{C}$ is the language of either UTVPIs
  or BUTVPIs, in polynomial time.
  %
  The \cplex{NP}-hardness of the consistency problem when $\lang{C}$ is the
  language of BUTVPIs follows by Lemma~\ref{lem:sat-to-consistency} from the
  fact that the satisfiability problem for BUTVPIs is \cplex{NP}-hard.

  We will show that the consistency problem is \cplex{NP}-hard when $\lang{C}$
  is the language of UTVPIs by a reduction from SAT. %
  Given an instance of SAT as a set $\Phi = \{ C_1, \dotsc, C_n \}$ of clauses
  over (possibly negated) literals $L_1, \dotsc, L_k$, we will construct a set
  $\IC$ of CDCs (whose consequents are UTVPIs) that is consistent if and only if
  $\Phi$ is satisfiable. %
  To this end, consider a relation symbol $R$ of arity $k + 1$, with the last
  position interpreted over the integers. %
  With each literal $L_i$ we associate the equality $x_i = a$, and with each
  clause $C_j$ we associate the CDC:
  \begin{equation}
    \label{eq:cdc-clause}
    \left[ \bigwedge_{L_i \in C_j}{(x_i \neq a)} \right] \land
    \left[ \bigwedge_{{\lnot L_i} \in C_j}{(x_i = a)} \right] \rightarrow
    y_1 > 0 \enspace .
  \end{equation}
  Then, let $\IC$ consist of $\top \rightarrow y_1 \le 0$ and the CDCs of the
  form \eqref{eq:cdc-clause} associated with each clause. %
  The propositional theory associated with $\IC$ is
  \begin{equation*}
    \pdc = \{ \top \rightarrow v_1 \} \cup %
    \{ P_j \rightarrow v_2 \mid  1 \le j \le n \} \enspace ,
  \end{equation*}
  where $v_1 \mapsto y_1 \le 0$, $v_2 \mapsto y_1 > 0$ and
  \begin{equation}
    \label{eq:ante-pj}
    P_j = %
    \Bigl( {\bigwedge_{L_i \in C_j}{\lnot p_i^a}} \Bigr) %
    \land %
    \Bigl( {\bigwedge_{{\lnot L_i} \in C_j}{p_i^a}} \Bigr) %
    \enspace .
  \end{equation}
  For every valuation $\alpha$ of $\pvar(\pdc)$, the $\alpha$-filtering
  $\pdc^\alpha$ of $\pdc$ is either $\{ y_1 \le 0 \}$ or $\{ y_1 \le 0, y_1 > 0
  \}$. %
  Since the latter set is unsatisfiable, by Theorem~\ref{thm:cdc-sat} $\IC$ is
  consistent if and only if there exists a valuation $\alpha$ such that, for
  every $j \in \{1, \dotsc, n \}$, $P_j$ does not hold true under $\alpha$.
  Clearly, to each valuation $\alpha$ of $\pvar(\pdc)$ there corresponds a
  valuation $\alpha'$ of $L_1, \dotsc, L_k$, and vice versa, such that
  $\alpha(p_i^a) = \vartrue$ if and only if $\alpha'(L_i) = \vartrue$; in turn,
  $P_j$ is true under $\alpha$ if and only if $C_j$ is false under $\alpha'$. %
  Thus, $\IC$ is consistent if and only if $\Phi$ is satisfiable. %
  Therefore, since the given reduction is obviously polynomial, the claim
  follows.\hfill\IEEEQED
\end{pf}



\section{Lossless Selections under CDCs}
\label{sec:losslessness}

The technique described in the previous section can be opportunely extended and
applied for checking whether a set of selection views of the form \eqref{eq:sel}
is lossless under CDCs, that is, whether every source relation satisfying the
given CDCs can be reconstructed by union from the fragments into which it is
decomposed by the given view definitions.

In this section, we will characterise lossless horizontal decomposition in terms
of unsatisfiability in \lang{C}, where \lang{C} is closed under negation. %
The \emph{losslessness problem} in \lang{C} is the decision problem that takes
as input a horizontal decomposition $\dec$ specified by selections of the form
\eqref{eq:sel} and a set $\IC$ of CDCs and answers the question: ``Is $\dec$
lossless under $\IC$?'' %
We will show that this problem is \cplex{co-NP}-complete when $\lang{C}$ is the
language of either UTVPIs or BUTVPIs. %
For these languages, our characterisation provides an exponential-time algorithm
for deciding the losslessness of $\dec$ under $\IC$, by means of a number of
unsatisfiability checks in $\lang{C}$ which is exponentially bounded by the size
of $\IC$.

By definition, a horizontal decomposition $\dec$ of $R$ into $V_1, \dotsc, V_n$
is lossless under a set $\IC$ of CDCs over $R$ if $R^I = {V_1}^I \cup \dotsb
\cup {V_n}^I$ for every model $I$ of $\IC \cup \dec$. %
As the extension of each view symbol is always included in the extension of $R$,
the problem is equivalent to checking that there is no model $I$ of $\IC \cup
\dec$ where a tuple $\overline{t} \in R^I$ does not belong to any ${V_i}^I$. %
In turn, this means that for each definition in $\dec$, which has the form
\eqref{eq:sel}, the values in $\overline{t}$ at non-interpreted positions do not
satisfy $\cond$, or the values in $\overline{t}$ at interpreted positions do not
satisfy the $\lang{C}$-formula $\sel$.

The formulae in $\dec$ apply to one tuple at a time and, as already observed in
Section~\ref{sec:consistency}, so do CDCs; therefore we can again focus on
single tuples. %
With each equality we associate, as before, a propositional variable whose
truth-value determines whether the equality is satisfied. %
Given a valuation $\alpha$, we consider the set consisting of
$\lang{C}$-formulae in the r.h.s.\ of all the CDCs that are applicable under
$\alpha$ and the \emph{negation} of the selection condition $\dc(\overline{y})$
of each view definition in $\dec$ whose selection condition
$\cond(\overline{x})$ is satisfied by $\alpha$. %
Then, checking losslessness is equivalent to checking that there exists no
valuation $\alpha$ for which the above set of $\lang{C}$-formulae is
satisfiable. %
Indeed, from such a valuation and the corresponding assignment of values from
$\idom$ satisfying the relevant $\lang{C}$-formulae, we can obtain a tuple that
provides a counterexample to losslessness.

Similarly to what we did in Section~\ref{sec:consistency} for sets of CDCs, we
build a propositional theory associated with a given horizontal decomposition.

\begin{definition}
  \label{def:prop-dec}
  Let $\dec = \{ \phi_1, \dotsc, \phi_n \}$ be a horizontal decomposition. %
  For each $\phi_i \in \dec$, which has the form \eqref{eq:sel}, we build
  \begin{equation}
    \prop(\phi_i) = \pcond \rightarrow \pidf'_i \enspace ,
  \end{equation} 
  in which $\pidf'_i$ is either a fresh propositional variable associated (by
  means of $\idf$) with the $\lang{C}$-formula $\sel(\overline{y})$, if any,
  occurring in $\phi_i$, or $\bot$ otherwise.\footnotemark\ %
  We denote $\{\, \prop(\phi) \mathrel{|} \phi \in \dec \,\}$ by $\psel$ and we
  call it the \emph{propositional theory associated with $\dec$}.
\end{definition} 

\footnotetext{This is because the constraints in $\dec$ may not specify a
  $\lang{C}$-formula.}

We consider the set $\var(\psel)$ of propositional variables occurring in
$\psel$ partitioned into $\pvar(\psel) = \{ \var(\pcond) \mid (\pcond
\rightarrow \pidf_i) \in \psel \}$ and $\cvar(\psel) = \var(\psel) \setminus
\pvar(\psel)$.

Given a set $\IC$ of CDCs over $R$ and a horizontal decomposition $\dec$ of $R$,
the propositional theory associated with $\IC \cup \dec$ is $\propth = \pdc \cup
\psel$, where $\pdc$ and $\psel$ are the propositional theories of
Definition~\ref{def:prop-IC} and Definition~\ref{def:prop-dec} associated with
$\IC$ and $\dec$, respectively.
The set $\var(\propth) = \var(\pdc) \cup \var(\psel)$ of propositional variables
occurring in $\propth$ is partitioned into $\pvar(\propth) = \pvar(\pdc) \cup
\pvar(\psel)$ and $\cvar(\propth) = \cvar(\pdc) \cup \cvar(\psel)$.

\begin{definition}
  Let $\IC$ be a set of CDCs over $R$ and let $\dec$ be a horizontal
  decomposition of $R$. %
  The \emph{auxiliary theory $\axioms$ for $\propth = {\pdc \cup \psel}$}
  consists of the propositional formulae in $\psel$ whose r.h.s.\ is $\bot$ and
  the axioms of unique value for $\pvar(\propth)$.
\end{definition} 

Observe that the above is a proper extension of Definition~\ref{def:aux-theory}:
whenever $\dec$ is empty, the auxiliary theory for $\propth$ coincides with the
auxiliary theory for $\pdc$.

\begin{example}
  \label{exa:utvpi-dec-prop}
  The propositional theory associated with $\dec$ of Example~\ref{exa:running}
  is $\psel = \{ \lnot p_2^a \land p_3^b \rightarrow \bot, \top \rightarrow
  \pidf'_2, \lnot p_3^b \rightarrow \bot \}$. %
  Let $\propth = \pdc \cup \psel$, where $\pdc$ is the propositional theory
  already given in Example~\ref{exa:utvpi-cdc-prop}. %
  The association between the propositional variables in $\cvar(\propth)$ and
  UTVPIs is $\idf$ of Example~\ref{exa:utvpi-cdc-prop} extended with $\pidf'_2
  \mapsto {y_2 < 4}$, and the auxiliary theory for $\propth$ is $\axioms = \{
  \lnot p_2^a \land p_3^b \rightarrow \bot, \lnot p_3^b \rightarrow \bot \}$.
\end{example} 

\begin{definition}
  Let $\dec$ be a horizontal decomposition, and let $\alpha$ be a valuation of
  $\pvar(\psel)$. %
  The \emph{$\alpha$-filtering of $\psel$} is the set
  \begin{equation}
    \label{eq:alpha-dec}
    \begin{aligned}
      \psel^\alpha = \{\, \lnot \idf(\pidf') \mathrel{|} {}&
      (\pcond \rightarrow \pidf') \in \psel,\\
      &\qquad\qquad\alpha(\pcond) = \vartrue,\; \pidf' \neq \bot \,\} \enspace ,
    \end{aligned}
  \end{equation} 
  consisting of the \emph{negation} of $\lang{C}$-formulae associated with
  propositional variables that occur in some propositional formula of $\psel$
  whose l.h.s.\ holds true under $\alpha$.
\end{definition} 

Observe that in \eqref{eq:alpha-dec}, differently from \eqref{eq:alpha-cdc},
$\lang{C}$-formulae are negated. %
This is because a counter-instance $I$ to losslessness is such that ${V_1}^I
\cup \dotsb \cup {V_n}^I = \emptyset$ and $R^I$ has only one tuple; therefore,
whenever the formula $\cond(\overline{x})$ in the selection that defines a view
symbol is satisfied by $I$, the $\lang{C}$-formula $\dc(\overline{y})$, if any,
is not. %
On the other hand, the $\lang{C}$-formula in the consequent of a CDC must hold
whenever the condition in the antecedent is satisfied.

For a valuation $\alpha$ of $\pvar(\propth)$, the $\alpha$-filtering of
$\propth$ is the set $\propth^\alpha = \pdc^\alpha \cup \psel^\alpha$, which, as
$\lang{C}$ is closed under negation, consists of $\lang{C}$-formulae.

The main result of this section is the following characterisation of lossless
horizontal decomposition in terms of unsatisfiability in $\lang{C}$.

\begin{theorem}
  \label{thm:losslessness}
  Let $\dec$ be a horizontal decomposition of $R$, let $\IC$ be a set of CDCs
  over $R$, and let $\axioms$ be the auxiliary theory for $\propth = \pdc \cup
  \psel$. %
  Then, $\dec$ is lossless under $\IC$ if and only if the $\alpha$-filtering
  $\propth^\alpha = {\pdc^\alpha \cup \psel^\alpha}$ of $\propth$ is
  unsatisfiable for every valuation $\alpha$ of $\pvar(\propth)$ satisfying
  $\axioms$.
\end{theorem} 

Whenever the satisfiability of $\lang{C}$-formulae is decidable,
Theorem~\ref{thm:losslessness} provides an algorithm for deciding whether a
given horizontal decomposition is lossless. %
We illustrate this in our running example with UTVPIs.

\begin{example}
  \label{exa:utvpi-dec-sat}
  Consider $\propth$ and $\axioms$ from Example~\ref{exa:utvpi-dec-prop}. %
  The only valuation of $\pvar(\propth)$ satisfying $\axioms$ is $\alpha = \{ \,
  p_2^a \mapsto \vartrue,\, p_3^b \mapsto \vartrue \}$, for which the
  $\alpha$-filtering of $\propth$ is
  \begin{equation*}
    \propth^\alpha = \underbrace{\bigl\{\; {y_1 + y_2 \le 5},\, {y_2 \ge 2},\, %
      {y_1 - y_2 \ge 0} \;\bigr\}}_{\pdc^\alpha}
    \cup \underbrace{\bigl\{\; y_2 \ge 4 \;\bigr\}}_{\psel^\alpha} .
  \end{equation*}
  Note that $y_2 \ge 4$ in $\psel^\alpha$ is $\lnot \idf(\pidf'_2)$, that is,
  the negation of $y_2 < 4$. %
  The set $\propth^\alpha = \pdc^\alpha \cup \psel^\alpha$ is unsatisfiable
  because from $y_1 + y_2 \le 5$ and $y_2 \ge 4$ we get $y_1 \le 1$, which
  together with $y_1 - y_2 \ge 0$ yields $y_2 \le 1$, in conflict with $y_2 \ge
  2$. %
  So, the horizontal decomposition $\dec$ is lossless under $\IC$.\footnotemark
\end{example} 

\footnotetext{%
  In the scenario of our running example it would makes sense to require
  salaries and bonuses to be non-negative quantities, which can be done by
  consistently adding the CDCs $\top \rightarrow y_1 \ge 0$ and $\top
  \rightarrow y_2 \ge 0$ without affecting the losslessness of the
  decomposition.}

We will now give the proof of Theorem~\ref{thm:losslessness}, for which we first
need to prove two additional lemmas. %
In the following, and in the rest of the article, let $\phitilde$ denote the
formula $\forall \overline{x},\overline{y} \mathrel{.}  R(\overline{x},
\overline{y}) \leftrightarrow \bigvee_{V \in \V}{V(\overline{x},\overline{y})}$,
and recall that a horizontal decomposition $\dec$ is lossless under $\IC$ if and
only if $\IC \cup \dec \models \phitilde$. %
We start by showing that, when $\IC$ consists of CDCs, $\IC \cup \dec$ does not
entail $\phitilde$ precisely if there is a counterexample to it with only one
tuple.

\begin{lemma}
  \label{lem:1-tuple-inst}
  Let $\dec$ be a horizontal decomposition of $R$ and let $\IC$ be a set of CDCs
  over $R$. %
  Then, $\IC \cup \dec \not\models \phitilde$ if and only if there exists a
  tuple $\overline{t}$ such that the instance $I = \{ R(\overline{t}) \}$ is a
  model of $\IC \cup \dec$.
\end{lemma}

\begin{pf}
  The ``if'' is trivial. %
  For the ``only if'', assume that $\IC \cup \dec \not\models \phitilde$. %
  Then, there exists a model $J$ of $\IC \cup \dec$ such that $J \not\models
  \phitilde$, that is, $R^J \neq {{V_1}^J \cup \dotsb \cup {V_n}^J}$. %
  The extension of each $V_i$ always contains a subset of the tuples in the
  extension of $R$ under every instance, hence there must be $\overline{t} \in
  R^J$ such that $\overline{t} \not\in {V_i}^J$ for every $i \in \{1, \dotsc,
  n\}$. %
  Let $I = \{ R(\overline{t}) \}$; as every constraint in $\IC \cup \dec$ is in
  one tuple and $J \models \IC \cup \dec$, we have that $I \models \IC \cup
  \dec$.\hfill\IEEEQED
\end{pf}

The next lemma is more technical: intuitively, it shows that any tuple that is
consistent with a valuation $\alpha$ satis\-fying the auxiliary theory provides a
counterexample to losslessness if and only if the assignment induced by its
interpreted positions satisfies the $\alpha$-filtering.

\begin{lemma}
  \label{lem:key-losslessness}
  Let $\dec$ be a horizontal decomposition of $R$, let $\IC$ be a set of CDCs
  over $R$, and let $\axioms$ be the auxiliary theory for $\propth = \pdc \cup
  \psel$. %
  Let $\alpha$ be a valuation of $\pvar(\propth)$, let $\overline{t}$ be a tuple
  consistent with $\alpha$, and let $\beta$ be the assignment induced by the
  interpreted positions of $\overline{t}$. %
  Whenever $\alpha \models \axioms$, we have that $\{ R(\overline{t}) \} \models
  {\IC \cup \dec}$ if and only if $\beta$ satisfies $\propth^\alpha$.
\end{lemma} 

\begin{pf}
  Let $n = \arity{R}$ and $k = \iattr{R}$.

  \begin{enumerate}[claim]
  \item Let $\prop(\phi) = \pcond \rightarrow \pidf'$, with $\phi \in \dec$. %
    Then, $I \not\models \phi$ iff $\alpha(\pcond) = \vartrue$ and, whenever
    $\pidf' \neq \bot$, $\beta \models \idf(\pidf')$.%
    \label{cla:key-lossless-1}
  \end{enumerate}

  \begin{description}[claimproof]
  \item[Proof.]  Since $\phi \in \dec$ has the form \eqref{eq:sel}, $I
    \not\models \phi$ iff $\cond(\overline{x})$ is true under $\{ x_1 \mapsto
    \overline{t}[1], \dotsc, x_k \mapsto \overline{t}[k] \}$ and
    $\sel(\overline{y})$, if any, holds true under $\{ y_{1} \mapsto
    \overline{t}[k+1], \dotsc, y_{n-k} \mapsto \overline{t}[n] \}$. %
    As $\pidf' \neq \bot$ iff $\phi$ contains a $\lang{C}$-formula
    $\sel(\overline{y}) = \idf(\pidf')$, the claim follows by construction of
    $\alpha$ and $\beta$.
  \end{description}

  Assume $\alpha \models \axioms$. %
  We will show that $I \not\models {\IC \cup \dec}$ if and only if $\beta$ does
  not satisfy $\propth^\alpha$.

  \begin{description}[iff]
  \item[``if''.] Assume $\beta \not\models \propth^\alpha$. %
    Then, there is a $\lang{C}$-formula $\psi \in \propth^\alpha$ that is not
    satisfied by $\beta$. %
    By construction of $\propth^\alpha$, either $\psi$ or its negation appear in
    some $\phi \in \IC \cup \dec$, depending on whether $\phi \in \IC$ or $\phi
    \in \dec$, respectively. %
    If $\phi \in \IC$, then $\prop(\phi) = \pcond \rightarrow \pidf$ with
    $\alpha(\pcond) = \vartrue$, so $I \not\models \phi$ (by
    \ref{cla:key-consist-2} in the proof of Lemma~\ref{lem:key-consistency}). %
    If $\phi \in \dec$, then $\prop(\phi) = \pcond \rightarrow \pidf'$ with
    $\pidf' \neq \bot$ and $\alpha(\pcond) = \vartrue$, hence $I \not\models
    \phi$ by \ref{cla:key-lossless-1}. %
    In either case $I \not\models \IC \cup \dec$.

  \item[``only if''.] Assume $I \not\models \IC \cup \dec$. %
    Then, there is some $\phi \in \IC \cup \dec$ that is not satisfied by $I$. %
    If $\phi$ is in $\IC$, by Lemma~\ref{lem:key-consistency} $\beta \not\models
    \pdc^\alpha$, hence $\beta \not\models \propth^\alpha$. %
    If $\phi$ is in $\dec$, $\prop(\phi) = \pcond \rightarrow \pidf'$; as $I
    \not\models \phi$, by \ref{cla:key-lossless-1} $\alpha(\pcond) = \vartrue$
    and $\pidf' \neq \bot$ implies $\beta \models \idf(\pidf)$. %
    Suppose $\pidf' = \bot$, then $\prop(\phi)$ is in $\axioms$ and, since
    $\alpha \models \axioms$, we obtain $\alpha(\pcond) = \varfalse$, which is a
    contradiction. %
    So, $\pidf' \neq \bot$ and $\beta \models \idf(\pidf')$. %
    In turn, we have that $\beta \not\models \lnot\idf(\pidf')$ and
    $\lnot\idf(\pidf') \in \propth^\alpha$. %
    Therefore, $\beta \not\models \propth^\alpha$.\hfill\IEEEQED
  \end{description}
\end{pf} 

\begin{thmpf}[of Theorem~\ref{thm:losslessness}]
  Let $n = \arity{R}$ and $k = \iattr{R}$. %
  We will show that $\IC \cup \dec \not\models \phitilde$ if and only if there
  exist $\alpha$ and $\beta$ satisfying $\axioms$ and $\propth^\alpha$,
  respectively. %

  \begin{description}[iff]
  \item [``if''.] Let $\alpha$ and $\beta$ be such that $\alpha \models\axioms$
    and $\beta \models \propth^\alpha$. %
    Since $\alpha \models \axioms$, no two distinct propositional variables in
    $\pvar(\propth)$ associated with the same position are both true under
    $\alpha$. %
    Hence, there is a tuple $\overline{t}$ consistent with $\alpha$ and such
    that $\beta$ is the assignment induced by its interpreted positions. %
    So, the instance $I = \{ R(\overline{t}) \}$ is a model of $\IC \cup \dec$
    by Lemma~\ref{lem:key-losslessness}. %
    Thus, as $I \not\models \phitilde$, ${\IC \cup \dec} \not\models
    \phitilde$ by Lemma~\ref{lem:1-tuple-inst}.

  \item[``only if''.] Assume that ${\IC \cup \dec} \not\models \phitilde$. %
    By Lemma~\ref{lem:1-tuple-inst}, there exists a tuple $\overline{t}$ such
    that $I = \{ R(\overline{t}) \}$ is a model of ${\IC \cup \dec}$. %
    Let $\beta$ be the assignment induced by the interpreted positions of
    $\overline{t}$, and let $\alpha$ be the valuation such that, for each $p \in
    \pvar(\propth)$, $\alpha(p) = \vartrue$ if $p = p_i^a$ and $\overline{t}[i]
    = a$, and $\alpha(p) = \varfalse$ otherwise.
    We will show that $\alpha$ satisfies $\axioms$ and, in turn, $\beta \models
    \propth^\alpha$ by Lemma~\ref{lem:key-losslessness}, since $I \models \IC
    \cup \dec$.
    By construction, $\alpha$ satisfies every propositional formula in $\axioms$
    of the form ${p_i^a \land p_i^b} \rightarrow \bot$ with $p_i^a,p_i^b \in
    \pvar(\propth)$ and $p_i^a \neq p_i^b$. %
    All other propositional formulae in $\axioms$ have the form $\prop(\phi) =
    \pcond \rightarrow \bot$, where $\phi$ is a constraint in $\dec$ that does
    not contain a $\lang{C}$-formula $\sel(\overline{y})$. %
    As $I \models \IC \cup \dec$, the condition $\cond(\overline{x})$ in each
    such $\phi \in \dec$ is not true under $\{ x_1 \mapsto \overline{t}[1],
    \dotsc, x_k \mapsto \overline{t}[k] \}$. %
    Therefore, $\prop(\phi) = \pcond \rightarrow \bot$ is true under $\alpha$ as
    $\alpha(\pcond) = \varfalse$ by construction of $\alpha$.\hfill\IEEEQED
  \end{description}
\end{thmpf} 

The unsatisfiability problem for $\lang{C}$ is the complement of the
satisfiability problem for $\lang{C}$.

\begin{lemma}
  \label{lem:unsat-to-losslessness}
  The unsatisfiability problem for $\lang{C}$ linearly reduces to the
  losslessness problem in $\lang{C}$.
\end{lemma} 

\begin{pf}
  Let $\Gamma = \{ \phi_1, \dotsc, \phi_n \}$ be a set $\lang{C}$-formulae. %
  We will show how to construct a horizontal decomposition that is lossless
  under $\IC = \emptyset$ precisely if $\Gamma$ is unsatisfiable. %
  To this end, take $\dec = \{\, V_i \colon \lnot\phi_i \mid \phi_i \in \Gamma
  \,\}$ and observe that, as $\lang{C}$ is closed under negation, $\lnot\phi_i
  \in \lang{C}$. %
  Thus, $\dec$ consists of selections of the form \eqref{eq:sel}, where $\sel =
  \lnot\phi_i$ and $\cond = \top$. %
  Therefore, $\dec$ is indeed a horizontal decomposition.

  Let $\propth = \pdc \cup \psel = \emptyset \cup \{\, \pidf'_i \mid i = 1,
  \dotsc, n \,\}$ for which $\idf = \{\, \pidf'_i \mapsto \lnot\phi_i \mid
  \phi_i \in \Gamma \,\}$. %
  Then, the auxiliary theory for $\propth$ is $\axioms = \emptyset$. %
  Since $\pvar(\propth) = \emptyset$, the only valuation of $\pvar(\propth)$ is
  $\alpha = \emptyset$, which satisfies $\axioms$ and for which $\propth^\alpha
  = \{\, \lnot\idf(\pidf'_i) \mid \pidf'_i \in \cvar(\psel) \,\} = \Gamma$. %
  Therefore, by Theorem~\ref{thm:losslessness}, $\dec$ is lossless under $\IC =
  \emptyset$ if and only if $\Gamma$ is unsatisfiable. %
  The reduction is linear in the size of $\Gamma$.\hfill\IEEEQED
\end{pf}

With regard to the losslessness problem in the languages of UTVPIs and BUTVPIs,
we have the following complexity results.

\begin{theorem}
  \label{thm:losslessness-probl}
  When $\lang{C}$ is the language of either BUTVPIs or UTVPIs, the losslessness
  problem in $\lang{C}$ is \cplex{co-NP}-complete.
\end{theorem} 

\begin{pf}
  Constructing the propositional theories $\propth$ and $\axioms$ takes linear
  time, checking whether a valuation $\alpha$ of $\pvar(\propth)$ satisfies
  $\axioms$ requires polynomial time, and checking that an assignment of
  integers to the variables in $\overline{y}$ satisfies $\propth^\alpha$ (whose
  construction takes linear time) can be done in polynomial time, whether
  $\propth^\alpha$ consists of UTVPIs or BUTVPIs. %
  Hence, in light of Theorem~\ref{thm:losslessness}, we can verify a given
  solution to the complement of the losslessness problem, in either language, in
  polynomial time. %
  Therefore, the losslessness problem is in \cplex{co-NP} in both cases.

  The \cplex{co-NP}-hardness in the case of BUTVPIs follows by
  Lemma~\ref{lem:unsat-to-losslessness} from the fact that the satisfiability
  problem for BUTVPI-formulae is \cplex{NP}-hard and so its complement is, in
  turn, \cplex{co-NP}-hard.

  We will show the \cplex{co-NP}-hardness of the losslessness problem when
  \lang{C} is the language of UTVPIs by a reduction from UNSAT. %
  The reduction is quite similar to the one given in the proof of
  Theorem~\ref{thm:consistency-probl} for showing the \cplex{NP}-hardness of the
  consistency problem for CDCs when \lang{C} is the language of UTVPIs. %
  Given a set $\Phi = \{ C_1, \dotsc, C_n \}$ of clauses over possibly negated
  literals $L_1, \dotsc, L_k$, we will build a set $\IC$ of CDCs (whose
  consequents are UTVPIs) and a horizontal decomposition $\dec$ (where the
  selection conditions on the interpreted positions are UTVPIs) such that $\dec$
  is lossless under $\IC$ if and only if $\Phi$ is unsatisfiable. %
  To this end, consider a source relation symbol $R$ of arity $k + 1$, with the
  last position interpreted over the integers, and the view symbol $V$. %
  With each literal $L_i$ we associate the equality $x_i = a$, and with each
  clause $C_j$ we associate the CDC \eqref{eq:cdc-clause}. %
  Let $\IC$ consist of the CDCs of the form \eqref{eq:cdc-clause} associated
  with each clause, and let $\dec$ be the horizontal decomposition specified by
  $V \colon y_1 > 0$.
  The propositional theory associated with $\IC \cup \dec$ is
  \begin{equation*}
    \propth = \{ P_j \rightarrow v \mid  1 \le j \le n \} %
    \cup \{ \top \rightarrow v' \} \enspace ,
  \end{equation*}
  with $v \mapsto y_1 > 0$, $v' \mapsto y_1 > 0$ and $P_j$ as in
  \eqref{eq:ante-pj}.
  For every valuation $\alpha$ of $\pvar(\propth)$, the $\alpha$-filtering
  $\propth^\alpha$ of $\propth$ is either $\{ y_1 > 0,\, y_1 \le 0 \}$ or $\{
  y_1 \le 0 \}$, depending on whether $\IC$ contains a CDC that is applicable
  under $\alpha$. %
  As the latter set is satisfiable, by Theorem~\ref{thm:losslessness} we get
  that $\dec$ is lossless under $\IC$ if and only if for every valuation
  $\alpha$ there exists $j \in \{1, \dotsc, n \}$ such that $P_j$ holds true
  under $\alpha$.
  Clearly, each valuation $\alpha$ of $\pvar(\propth)$ corresponds to a
  valuation $\alpha'$ of $L_1, \dotsc, L_k$, and vice versa, such that
  $\alpha(p_i^a) = \vartrue$ if and only if $\alpha'(L_i) = \vartrue$; in turn,
  $P_j$ is true under $\alpha$ if and only if $C_j$ is false under $\alpha'$. %
  Thus, $\dec$ is lossless under $\IC$ if and only if $\Phi$ is unsatisfiable. %
  Therefore, since the given reduction is obviously polynomial, the claim
  follows.\hfill\IEEEQED
\end{pf}



\section{Adding FDs and UINDs}
\label{sec:separability}

So far, we have considered lossless horizontal decompo\-sition under CDCs in
isolation; in this section, we extend our study to the case in which the
integrity constraints over the source schema are combinations of CDCs with more
traditional database constraints. %
This investigation is vital to understand whether, how and to what extent the
techniques we described in Section~\ref{sec:losslessness} can be applied to
existing database schemas on which a set of integrity constraints other than
CDCs is already defined.

Here, we focus on two well-known classes of integrity constraints, namely
functional dependencies (FDs) and unary inclusion dependencies (UINDs)
\cite{Abiteboul:1995:FD}. %
Under certain restrictions -- as we shall see -- their interaction with CDCs can
be fully captured, w.r.t.\ lossless horizontal decomposition, in terms of
CDCs. %
It is important to remark that we consider restrictions solely on the CDCs, so
that existing integrity constraints need not be modified in any way in order to
allow for CDCs.

Let us recall that an instance $I$ satisfies a UIND $R[i] \subseteq R[j]$ if
every value in the $i$-th column of $R^I$ appears in the $j$-th column of
$R^I$. %
The following example shows that, if we allow CDCs together with constraints
from another class, such as UINDs, their interaction may influence the
losslessness of horizontal decomposition.

\begin{example}
  \label{exa:cdc-uind}
  Let $R$ and $V$ be relation symbols of arity $2$, whose positions are
  interpreted over the integers. %
  Let $\dec$ be the horizontal decomposition defined by $V \colon y_1 > 3$, and
  let $\IC$ be a set of integrity constraints on $R$ consisting of the CDC $\top
  \rightarrow y_2 > 3$ and the UIND $R[1] \subseteq R[2]$. %
  It is easy to see that $\IC$ entails $\top \rightarrow y_1 > 3$. %
  Therefore, $\dec$ is lossless under $\IC$ because $V$ selects all of the
  tuples in $R$, which is clearly not the case without the UIND.
\end{example} 

We now introduce a general property, \emph{separability}, that will constitute
the main technical tool for the subsequent analysis of combinations of CDCs with
FDs and UINDs.
Informally, a class of constraints is separable from CDCs if, after
making explicit the result of their interaction, which is captured by a suitable
set of inference rules, we can disregard constraints from that class and focus
solely on CDCs, as far as lossless horizontal decomposition is concerned.

In what follows, for a set $\IC$ of constraints we denote by $\cdc(\IC)$ the
maximal subset of $\IC$ consisting solely of CDCs.

\begin{definition}[Separability]
  \label{def:separability}
  Let $\cls{C}$ be a class of integrity constraints, let $\rules{S}$ be a finite
  set of sound inference rules\footnote{We assume the reader to be familiar with
    the standard notions (from proof theory) of inference rule, soundness,
    deductive closure.}  for $\cls{C}$ extended with CDCs, and let $\IC$ consist
  of CDCs and $\cls{C}$-constraints. %
  We say that the $\cls{C}$-constraints are \emph{$\rules{S}$-separable} in
  $\IC$ from the CDCs if every horizontal decomposition is lossless under $\IC$
  exactly when it is lossless under $\cdc(\IC^*)$, where $\IC^*$ denotes the
  $\rules{S}$-closure of $\IC$.%
  \footnote{As the constraints that are not CDCs are in any case filtered out
    from $\IC^*$, it does not matter whether $\cls{C}$ extended with CDCs is
    closed under $\rules{S}$ or not.}  %
  We say that the $\cls{C}$-constraints are \emph{separable} if there is some
  $\rules{S}$ for which they are $\rules{S}$-separable.
\end{definition} 

Thus, to check whether a horizontal decomposition $\dec$ is lossless under an
$\rules{S}$-separable combination $\IC$ of CDCs and other constraints, one can
proceed as follows:
\begin{enumerate}
\item compute the deductive closure $\IC^*$ of $\IC$ w.r.t.\ $\rules{S}$, which
  makes explicit the interaction between CDCs and the other constraints in $\IC$
  by adding entailed constraints;
\item by using the technique of Section~\ref{sec:losslessness}, check whether
  $\dec$ is lossless under $\cdc(\IC^*)$, that is, the set obtained by
  discarding from $\IC^*$ all of the constraints that are not CDCs.
\end{enumerate}

Observe that $\rules{S}$-separability implies $\rules{S}'$-separability for
every sound $\rules{S}' \supseteq \rules{S}$.

\subsection{Functional Dependencies}

We begin our investigation of separability by showing that FDs do not interact
with CDCs and so, as far as the losslessness of horizontal decomposition is
concerned, they can be freely allowed in combination with them.

\begin{theorem}
  \label{thm:adding-fds}
  Let $\IC$ be a set of CDCs and FDs. %
  Then, the FDs are $\emptyset$-separable in $\IC$ from the CDCs.
\end{theorem} 

\begin{pf}
  We will prove that a horizontal decomposition is lossless under $\IC$ if and
  only if it is lossless under $\cdc(\IC)$.

  \begin{description}[iff]
  \item[``if''.] We have that $\cdc(\IC) \subseteq \IC$ and, in turn, $\IC$
    entails $\cdc(\IC)$; therefore $\cdc(\IC) \models \phitilde$ implies $\IC
    \models \phitilde$.\footnote{Recall that $\phitilde = \forall
      \overline{x},\overline{y} \mathrel{.}  R(\overline{x}, \overline{y})
      \leftrightarrow \bigvee_{i=1}^n{V_i(\overline{x},\overline{y})}$.}
  \item[``only if''.] %
    Whenever a horizontal decomposition is not lossless, by
    Lemma~\ref{lem:1-tuple-inst} there is a witness instance $I$ with only one
    tuple. %
    Since the violation of an FD involves at least two tuples, $I$ satisfies all
    of the FDs in $\IC$.\footnote{As a matter of fact, it satisfies any FD.}
    \hfill\IEEEQED
  \end{description}
\end{pf}

\subsection{Unary Inclusion Dependencies}

Since in general it is not possible to compare values from $\dom$ with values
from $\idom$, we consider only UINDs of the form $R[i] \subseteq R[j]$ where
positions $i$ and $j$ are either both non-interpreted or both interpreted. %
We refer to the UINDs in the former case as \emph{X-UINDs} and in the latter as
\emph{Y-UINDs}.
Let $n = \arity{R}$ and $k = \iattr{R}$; we write $R[x_i] \subseteq R[x_j]$ with
$i,j \in \{ 1, \dotsc, k \}$ to denote the X-UIND $R[i] \subseteq R[j]$ and we
write $R[y_i] \subseteq R[y_j]$ with $i,j \in \{ 1, \dotsc, n-k \}$ to denote
the Y-UIND $R[i+k] \subseteq R[j+k]$.

\subsubsection*{UINDs on Interpreted Attributes}

First, we study the interaction between Y-UINDs (that is, UINDs at interpreted
positions) and a restricted form of CDCs, which we shall introduce shortly. %
This interaction is captured by the following \emph{domain propagation} rule:
\begin{equation}
  \frac{%
    {\top \rightarrow \dc(y_i)} \qquad \qquad {R[y_j] \subseteq R[y_i]}
  }{\top \rightarrow \dc(y_j)} \enspace ,
  \tag{\text{dp}}
  \label{eq:dom-prop}
\end{equation} 
whose soundness is easily shown below.

\begin{theorem}
  \label{thm:sound-dp}
  Let $\IC$ be a set of CDCs and UINDs. %
  If $\IC \models \forall \overline{x}, \overline{y} \,.\, R(\overline{x},
  \overline{y}) \rightarrow \dc(y_i)$ and $\IC \models {R[y_j] \subseteq
    R[y_i]}$, then $\IC \models \forall \overline{x}, \overline{y} \,.\,
  R(\overline{x}, \overline{y}) \rightarrow \dc(y_j)$.
\end{theorem} 

\begin{pf}
  If $\IC$ is inconsistent, the claim follows trivially. %
  Thus, let $I$ be a model of $\IC$; hence $I$ satisfies the CDC $\forall
  \overline{x}, \overline{y} \mathrel{.} R(\overline{x}, \overline{y})
  \rightarrow \dc(y_i)$ and the UIND $R[y_j] \subseteq R[y_i]$. %
  If $R^I = \emptyset$, then trivially $I \models \forall \overline{x},
  \overline{y} \mathrel{.} R(\overline{x},\overline{y}) \rightarrow \dc(y_j)$. %
  So, let $R^I \neq \emptyset$ and suppose $I \not\models \forall \overline{x},
  \overline{y} \mathrel{.} R(\overline{x}, \overline{y}) \rightarrow
  \dc(y_j)$. %
  Then, there exists $\overline{t} \in R^I$ for which $\dc(\overline{t}[j+k])$
  holds true, with $k = \iattr{R}$. %
  By the UIND, there must be $\overline{t}' \in R^I$ such that
  $\overline{t}'[i+k] = \overline{t}[j+k]$. %
  Hence $\dc(\overline{t}'[i+k])$ is not true, in contradiction of $I \models
  \forall \overline{x}, \overline{y} \mathrel{.} R(\overline{x}, \overline{y})
  \rightarrow \dc(y_i)$.\hfill\IEEEQED
\end{pf}

It turns out that when all of the CDCs that mention a variable $y$ corresponding
to an interpreted position affected by some Y-UIND have the form $\top
\rightarrow \dc(y)$, the domain propagation rule fully captures the interaction
between such CDCs and Y-UINDs w.r.t.\ losslessness.

\begin{definition}
  \label{def:dp-controllable}
  We say that a set $\IC$ of CDCs and Y-UINDs is \emph{dp-controllable} if, for
  every Y-UIND $R[y_i] \subseteq R[y_j]$ in $\IC$ with $i \neq j$, all of the
  CDCs in $\IC$ mentioning the variable $y$, where $y$ is $y_i$ or $y_j$, are of
  the form $\top \rightarrow \dc(y)$.
\end{definition} 

\begin{theorem}
  \label{thm:yuind-sep}
  Let $\IC$ be a dp-controllable set of CDCs and Y-UINDs. %
  Then, the Y-UINDs are $\{ \eqref{eq:dom-prop} \}$-separable in $\IC$ from the
  CDCs.\hfill\IEEEQED
\end{theorem} 

The above theorem is a special case of a more general result
(Theorem~\ref{thm:uind-sep}) given later on.

Even though in general dp-controllability is not a necessary condition for the
$\{\eqref{eq:dom-prop}\}$-separability of Y-UINDs from CDCs, the following
examples show two different situations where, in the absence of
dp-controllability, the Y-UINDs are not $\{\eqref{eq:dom-prop}\}$-separable from
the CDCs.

\begin{example}
  Let $R$ be a ternary relation symbol, whose last two positions are interpreted
  over the integers. %
  Let $\IC$ consist of the Y-UIND $R[y_1] \subseteq R[y_2]$ and of the CDCs
  ${x_1 = a} \rightarrow {y_2 > 2}$, ${x_1 \neq a} \rightarrow {y_1 > 0}$, ${x_1
    \neq a} \rightarrow {y_1 < 0}$, and consider the view symbol $V \colon y_1 >
  1$. %
  For $x_1 \neq a$ there is no suitable value for $y_1$ to satisfy the above
  CDCs, thus every model $I$ of $\IC$ is such that, for every $\overline{t} \in
  R^I$, $\overline{t}[1] = a$ and $\overline{t}[3] > 2$. %
  Moreover, by the Y-UIND $R[y_1] \subseteq R[y_2]$, we also have that
  $\overline{t}[2] > 2$, and therefore every tuple in $R^I$ is also in $V^I$,
  which means that $V$ is lossless under $\IC$. %
  Clearly, this is not the case in the absence of the Y-UIND, that is, under
  $\cdc(\IC)$. %
  Let $\cl$ be the $\{\eqref{eq:dom-prop}\}$-closure of $\IC$. %
  Then, as $\cl = \IC$, we have that $V$ is lossy under $\cdc(\cl)$ and,
  therefore, the Y-UIND is not $\{\eqref{eq:dom-prop}\}$-separable in $\IC$ from
  the CDCs.
\end{example}

\begin{example}
  Let $R$ be a relation symbol of arity $4$ and with all of its positions
  interpreted over the integers. %
  Consider the view symbol $V \colon {y_3 < 3} \land {y_4 > 4}$, and let $\IC$
  consist of the Y-UIND $R[y_1] \subseteq R[y_2]$ and the CDCs $\top \rightarrow
  {y_1 + y_3 > 0}$, $\top \rightarrow {y_2 + y_4 \le 0}$, $\top \rightarrow {y_3
    - y_4 \le 0}$. %
  The above CDCs entail $\top \rightarrow {y_1 - y_2 \ge 1}$, thus in every
  model $I$ of $\IC$ each tuple $\overline{t} \in R^I$ must be such that
  $\overline{t}[1] - \overline{t}[2] \ge 1$.

  By the Y-UIND $R[y_1] \subseteq R[y_2]$, for each $d$ in
  $\pi_2(R^I)$\footnote{$\pi_i$ denotes projection on the $i$-th position.}
  there exists $d' \in \pi_2(R^I)$ with $d' \ge d+1$. %
  Then, as $d' \neq d$, the instance $I$ is either infinite or empty. %
  Hence, every horizontal decomposition is lossless under $\IC$.

  On the other hand, let $\cl$ be the $\{\eqref{eq:dom-prop}\}$-closure of $\IC$
  and observe that $\cl = \IC$. %
  Let $J = \{ R(1,0,0,0) \}$; then, since $J \models \cdc(\cl)$, $V$ is lossy
  under $\cdc(\cl)$. %
  Therefore, the Y-UIND is not $\{\eqref{eq:dom-prop}\}$-separable in $\IC$ from
  the CDCs.
\end{example}

\subsubsection*{UINDs on Non-Interpreted Attributes}

We now turn our attention to combinations of CDCs and X-UINDs (i.e., UINDs at
non-interpreted positions). %
First, we show that the syntactic restrictions introduced in
\cite{FeinererFG:2013:bncod} on the CDCs are \emph{not} sufficient for the
$\emptyset$-separability of X-UINDs. %
Indeed, the following is a counterexample to Theorem~7 of
\cite{FeinererFG:2013:bncod}.

\begin{example}
  \label{exa:non-sep}
  Let $R$ be a ternary relation symbol, with the third position interpreted over
  the integers. %
  Let $\IC$ consist of the CDC ${x_2 = a} \rightarrow {y_1 \le 0} \land {y_1 >
    0}$ and the X-UIND $R[1] \subseteq R[2]$. %
  The CDCs in $\IC$ are trivially \emph{non-overlapping} with the UINDs
  \cite{FeinererFG:2013:bncod} and \emph{partition-free}
  \cite{FeinererFG:2013:bncod}.
  Consider the horizontal decomposition $\dec$ specified by the selections $V_1
  \colon x_1 \neq a$, $V_2 \colon x_2 \neq b$ and $V_3 \colon y_1 \neq 0$. %
  Observe that every tuple other than $(a,b,0)$ is captured by at least one of
  the above selections. %
  Let $I = \{ R(a,b,0) \}$; clearly, $I \models \cdc(\IC) \cup \dec$ but $I
  \not\models \phitilde$, hence $\dec$ is not lossless under $\cdc(\IC)$.
  However, $\dec$ is lossless under $\IC$ as every model of $\IC \cup \dec$ also
  satisfies $\phitilde$. %
  This is due to the fact that there exists no instance $J$ such that $J \models
  \IC$ and $R(a,b,0) \in J$. %
  Indeed, to satisfy the UIND $R[1] \subseteq R[2]$, such an instance $J$ must
  also contain a tuple $\overline{t} \in R^J$ with $\overline{t}[2] = a$ which,
  on the other hand, does not satisfy the CDC ${x_2 = a} \rightarrow {y_1 \le 0}
  \land {y_1 > 0}$. %
  Hence, the X-UINDs are not $\emptyset$-separable in $\IC$ from the CDCs.
\end{example} 

Below, we introduce a restriction on the CDCs, which ensures the
$\emptyset$-separability of the X-UINDs.

\begin{definition}
  \label{def:global-consistency}
  A set $\IC$ of CDCs is \emph{globally consistent} if, for every
  $\iattr{R}$-tuple $\overline{t}_x$ of $\dom$ constants, there is a tuple
  $\overline{t}_y$ of $\arity{R}-\iattr{R}$ values from $\idom$ such that the
  instance $\{ R(\overline{t}) \}$, with $\overline{t} = \tuple{ \overline{t}_x,
    \overline{t}_y }$, is a model of $\IC$.
\end{definition} 

\noindent Note that $\cdc(\IC)$ in Example~\ref{exa:non-sep} is not globally
consistent.

\begin{theorem}
  \label{thm:xuind-sep}
  Let $\IC$ consist of CDCs and X-UINDs such that $\cdc(\IC)$ is globally
  consistent. %
  Then, the X-UINDs are $\emptyset$-separable in $\IC$ from the
  CDCs.\hfill\IEEEQED
\end{theorem} 

The above theorem, like Theorem~\ref{thm:yuind-sep}, is a special case of a more
general result (Theorem~\ref{thm:uind-sep}) given later on.

It is possible to check for the global consistency of a set $\IC$ of CDCs in a
way similar to the one described in Theorem~\ref{thm:cdc-sat} for consistency,
by building the propositional theory $\pdc$ associated with $\IC$, along with
the auxiliary theory $\axioms$ for $\pdc$, and then checking that the
$\alpha$-filtering $\pdc^\alpha$ of $\pdc$ is satisfiable \emph{for every}
(rather than just for one) valuation $\alpha$ of $\pvar(\pdc)$ that satisfies
$\axioms$. %
Indeed, under the assumptions of Theorem~\ref{thm:cdc-sat}, $\IC$ is globally
consistent if and only if $\pdc^\alpha$ is satisfiable for every valuation
$\alpha$ of $\pvar(\pdc)$ satisfying $\axioms$.
Checking for global consistency is expensive, because it requires an exponential
number of satisfiability checks in $\lang{C}$; the associated decision problem
is in \cplex{PSPACE} (the space used for one satisfiability check can be reused
for the next) for UTVPIs as well as for BUTVPIs.

Devising purely syntactic restrictions that guarantee the global consistency of
CDCs depends on the specific constraint language $\lang{C}$ in use, which is
indeed what we overlooked in \cite{FeinererFG:2013:bncod}. %
As it turns out, the non-overlapping and partition-free restrictions of
\cite{FeinererFG:2013:bncod} ensure global consistency (and so also the
$\emptyset$-separability of X-UINDs) only for sets of CDCs whose consequents are
UTVPIs. %
This is not the case anymore for CDCs whose consequents are BUTVPIs, which
indeed allow to express Example~\ref{exa:non-sep}.

We provide a condition that, although not guaranteeing global consistency,
ensures the $\emptyset$-separability of the X-UINDs. %
Moreover, this restriction can be checked more efficiently than global
consistency, as it requires only a polynomial number of
$\lang{C}$-satisfiability checks.

\begin{definition}
  \label{def:disjoint-cdcs}
  Let $\IC$ be a set of CDCs. %
  We say that the CDCs in $\IC$ are \emph{disjoint} w.r.t.\ an X-UIND
  $R[x_i] \subseteq R[x_j]$ if for any two distinct CDCs\footnote{Mistakenly, in
    \cite{Feinerer:2015:LSV} the CDCs were not required to be distinct.}
  $\phi_1(\overline{x}_1,\overline{y}_1)$ and
  $\phi_2(\overline{x}_2,\overline{y}_2)$ in $\IC$, with $x_j$ in
  $\overline{x}_1$, the consequent of $\phi_1$ is satisfiable and has no
  variables in common with the consequent of $\phi_2$.
\end{definition} 

Intuitively, the above requires that all of the variables appearing in the
consequent of any CDC $\phi$ whose antecedent mentions the variable $x_j$
affected by an X-UIND $R[x_i] \subseteq R[x_j]$ do not occur in the consequent
of any other CDC; moreover, the consequent of each such $\phi$ must be
satisfiable.

\begin{theorem}
  \label{thm:xuind-sep-disj}
  Let $\IC$ be a set of CDCs and X-UINDs, where the CDCs are disjoint w.r.t.\
  each X-UIND in $\IC$. %
  Then, the X-UINDs are $\emptyset$-separable in $\IC$ from the
  CDCs.\hfill\IEEEQED
\end{theorem} 

The above theorem is a special case of a more general result
(Theorem~\ref{thm:uind-sep-disj}) given later on in Section~\ref{sec:uinds-fds}.

Clearly, as global consistency is a property of the CDCs in isolation, whereas
disjointness is relative to a X-UIND, these two notions are incomparable, in the
sense that one does not imply the other and vice versa, as shown below.

\begin{example}
  \label{exa:incomp-glob-disj}
  Let $R$ be a ternary relation symbol, whose third position is interpreted over
  the integers, and let $\psi$ be the X-UIND $R[1] \subseteq R[2]$.
  The set $\IC_1$ consisting of the CDCs ${x_1 = a} \rightarrow {y_1 < 0}$ and
  ${x_1 = a} \rightarrow {y_1 > 0}$ is not globally consistent, as there is no
  suitable value for the third position (associated with $y_1$) whenever the
  value of the first (associated with $x_1$) is $a$; however, the CDCs in
  $\IC_1$ are disjoint w.r.t.\ $\psi$, since neither CDC mentions the variable
  $x_2$ affected by $\psi$.
  On the other hand, the set $\IC_2$ consisting of
  ${x_1 = a} \rightarrow {y_1 > 0}$ and ${x_2 = a} \rightarrow {y_1 > 1}$ is
  globally consistent, but it is not disjoint w.r.t.\ $\psi$, because the second
  CDC mentions $x_2$ in its antecedent, and the variable $y_1$ mentioned in its
  consequent also appears in the consequent of the first CDC.\footnote{The
    example given in \cite{Feinerer:2015:LSV} is incorrect.}
\end{example}

\subsubsection*{UINDs on All Attributes}

We now study the separability of UINDs (i.e., X-UINDs and Y-UINDs
together)\footnote{Recall that UINDs between non-interpreted and interpreted
  positions are not allowed, as they make little sense.} from CDCs.
The following is a generalisation of both Theorem~\ref{thm:yuind-sep} and
Theorem~\ref{thm:xuind-sep}.

\begin{theorem}
  \label{thm:uind-sep}
  Let $\IC$ be a set of globally consistent CDCs, X-UINDs and Y-UINDs, such that
  the CDCs and Y-UINDs are dp-controllable. %
  Then, the UINDs are $\{\eqref{eq:dom-prop}\}$-separable in $\IC$ from the
  CDCs.
\end{theorem} 

To give the proof of the above theorem, we will need to prove several lemmas,
showing how any given model of the (saturated set of) CDCs can be extended in
order to satisfy the UINDs as well.

\begin{lemma}
  \label{lem:same-value}
  Let $\IC$ be a dp-controllable set of CDCs and Y-UINDs, and let $\overline{t}$
  be a tuple such that $\{R(\overline{t})\} \models \cdc(\cl)$, where $\cl$ is
  the $\{\eqref{eq:dom-prop}\}$-closure of $\IC$. %
  Let $\psi = R[i] \subseteq R[j]$ be a Y-UIND in $\IC$, and let $\overline{t}'$
  be identical to $\overline{t}$ except for $\overline{t}'[j] =
  \overline{t}[i]$. %
  Then, $\{R(\overline{t}')\} \models \cdc(\cl) \cup \{ \psi \}$.
\end{lemma} 

\begin{pf}
  Let $\IC' = \cdc(\cl)$. %
  Since all of the UINDs in $\IC$ are Y-UINDs, $i,j > k$ with $k = \iattr{R}$. %
  As $\overline{t}$ satisfies $\IC'$ and $\overline{t}'$ differs from
  $\overline{t}$ only on the $j$-th element, $\overline{t}'$ satisfies every CDC
  in $\IC'$ not mentioning the variable $y_{j-k}$. %
  The only CDCs in $\IC'$ which are allowed to mention $y_{j-k}$ have the form
  $\top \rightarrow \dc(y_{j-k})$. %
  For each such CDC, since $R[i] \subseteq R[j]$ is in $\IC'$, by
  \eqref{eq:dom-prop} also $\top \rightarrow \dc(y_{i-k})$ is in $\IC'$. %
  Hence $\dc(\overline{t}[i])$ holds true, and in turn $\dc(\overline{t}'[j])$
  is true as well, because $\overline{t}'[j] = \overline{t}[i]$. %
  Therefore, $\overline{t}'$ satisfies all the CDCs of the form $\top
  \rightarrow \dc(y_{j-k})$. %
  Moreover, $\overline{t}'$ trivially satisfies the UIND $\psi$, as
  $\overline{t}'[i] = \overline{t}'[j] = \overline{t}[i]$.\hfill\IEEEQED
\end{pf}

\begin{lemma}
  \label{lem:model-yuinds}
  Let $\IC$ be a dp-controllable set of CDCs and Y-UINDs, and let $I$ be a model
  of $\cdc(\cl)$, where $\cl$ is the $\{\eqref{eq:dom-prop}\}$-closure of
  $\IC$. %
  Then, there exists an instance $J \supseteq I$ such that $J \models \cl$.
\end{lemma} 

\begin{pf}
  Let $J_0 = I$. %
  We will iteratively add tuples to $J_0$ so as to obtain a model of $\cl$. %
  At each iteration $k$, proceed as follows:
  \begin{enumerate}
  \item Find a violation of some Y-UIND $R[i] \subseteq R[j]$ in $\cl$, that is,
    a value $d \in \pi_i(R^{J_k})$ which is not in $\pi_j(R^{J_k})$.
  \item Take $\overline{t} \in R^{J_k}$ such that $\overline{t}[i] = d$ and
    $\overline{t}[j] \neq d$.
  \item Let $J_{k+1} = J_{k} \cup \{ R(\overline{t}') \}$, with $\overline{t}'$
    identical to $\overline{t}$ except for $\overline{t}'[j] = d$.
  \end{enumerate}

  \noindent In the worst case, to satisfy all the Y-UINDs, for every pair of
  interpreted positions $p$ and $q$ the above procedure will have to make the
  projections on $p$ and $q$ equal. %
  This is possible because, after each iteration $k$, $\adom(J_{k+1}) =
  \adom(J_k)$,\footnote{As a matter of fact, $\adom(J_{k+1}) \cap \idom =
    \adom(J_k) \cap \idom$ suffices.} as $\overline{t}'$ does not introduce new
  constants from $\dom$ or $\idom$. %
  In such a worst-case scenario, for each tuple $\overline{t} \in R^I$ and every
  interpreted position $p$, the value $\overline{t}[p]$ will be copied to every
  interpreted position other than $p$, resulting in the insertion of $r-1$ new
  tuples, with $r = \arity{R}-\iattr{R}$ (i.e., the number of interpreted
  positions in $R$). %
  The total number of tuples added to $I$ equals at most $m \cdot r \cdot
  (r-1)$, where $m$ is the number of tuples in $I$, and therefore the procedure
  terminates after finitely many steps, yielding an instance $J \supseteq I$
  that satisfies all the Y-UINDs in $\cl$ by construction.


  Let $\IC' = \cdc(\cl)$. %
  To conclude the proof, we show by induction that $J$ satisfies $\IC'$. %
  The base case is $J_0 \models \IC'$. %
  Observe that $\{ R(\overline{t}) \} \models \IC'$, because $\IC'$ consists of
  CDCs. %
  Then, assuming $J_k \models \IC'$, we have that $J_{k+1} \models \IC'$, since
  $\{ R(\overline{t}') \} \models \IC'$ by
  Lemma~\ref{lem:same-value}.\hfill\IEEEQED
\end{pf}

\begin{lemma}
  \label{lem:model-xuinds}
  Let $\IC$ consist of X-UINDs and globally consistent CDCs, and let $I$ be a
  model of $\cdc(\IC)$. %
  Then, there exists an instance $J$ such that $J \supseteq I$ and $J \models
  \IC$.
\end{lemma} 

\begin{pf}
  Let $\IC' = \cdc(\IC)$ and $J_0 = I$. %
  We will show how to build a model $J \supseteq I$ of $\IC$ by iteratively
  adding tuples to $J_0$. %
  At each iteration $k$, proceed as follows:
  \begin{enumerate}
  \item Find a violation of some X-UIND $R[i] \subseteq R[j]$ in $\IC$, i.e., a
    $\dom$ constant $a \in \pi_i(R^{J_k})$ which is not in $\pi_j(R^{J_k})$.
  \item Take $\overline{t} \in R^{J_k}$ such that $\overline{t}[i] = a$ and
    $\overline{t}[j] \neq a$.
  \item Let $J_{k+1} = J_{k} \cup \{ R(\overline{t}') \}$, where $\overline{t}'$
    agrees with $\overline{t}$ at non-interpreted positions except for
    $\overline{t}'[j] = a$. %
    Suitable values of $\overline{t}'$ at interpreted positions exist by
    Definition~\ref{def:global-consistency} as $\IC'$ is globally consistent.
  \end{enumerate}

  \noindent In the worst case, to satisfy all the X-UINDs, for each pair of
  non-interpreted positions $p$ and $q$ the above procedure will have to make
  the projections on $p$ and $q$ equal. %
  This is possible because, after each iteration $k$, $\adom(J_{k+1}) \cap \dom
  = \adom(J_k) \cap \dom$, as $\overline{t}'$ does not contain any new constant
  from $\dom$ (though it may contain new values from $\idom$). %
  In this worst-case scenario, for each tuple $\overline{t} \in I$ and every
  non-interpreted position $p$, the value $\overline{t}[p]$ will be copied to
  every non-interpreted position other than $p$, resulting in the insertion of
  $\iattr{R}$ new tuples. %
  The total number of tuples added to $I$ is at most equal to $m \cdot \iattr{R}
  \cdot (\iattr{R}-1)$, where $m$ is the number of tuples in $I$, and therefore
  the procedure terminates after finitely many steps, yielding an instance $J
  \supseteq I$ that satisfies all the X-UINDs in $\IC$ by construction.

  To conclude the proof, we show by induction that $J$ is a model $\IC'$. %
  The base case is $J_0 \models \IC'$. %
  Observe that $\{ R(\overline{t}) \} \models \IC'$, since $\IC'$ consists of
  CDCs. %
  Then, assuming $J_k \models \IC'$, we have that $J_{k+1} \models \IC'$ as $\{
  R(\overline{t}') \} \models \IC'$ by the global consistency of
  $\IC'$.\hfill\IEEEQED
\end{pf}

\begin{thmpf}[of Theorem~\ref{thm:uind-sep}]
  Let $\cl$ be the closure of $\IC$ under $\{\eqref{eq:dom-prop}\}$, and let
  $\IC' = \cdc(\cl)$. %
  Observe that $\cl \equiv \IC$, as \eqref{eq:dom-prop} is sound by
  Theorem~\ref{thm:sound-dp}. %
  According to Definition~\ref{def:separability}, we will show that $\IC \cup
  \dec \models \phitilde$ if and only if $\IC' \cup \dec \models \phitilde$.

  \begin{description}[iff]
  \item[``if''.] As $\IC' \cup \dec \subseteq \cl \cup \dec$, every model of
    $\cl \cup \dec$ is also a model of $\IC' \cup \dec$. %
    Hence, $\cl \cup \dec \models \IC' \cup \dec$ and, since $\cl \equiv \IC$,
    in turn $\IC \cup \dec \models \IC' \cup \dec$. %
    Therefore, $\IC \cup \dec \models \phitilde$ whenever $\IC' \cup \dec
    \models \phitilde$.

  \item[``only if''.] By contraposition. %
    Assume that ${\IC' \cup \dec} \not\models \phitilde$. %
    Then, as $\IC'$ consists solely of CDCs, by Lemma~\ref{lem:1-tuple-inst}
    there is a tuple $\overline{t}$ such that $I = \{ R(\overline{t}) \}$
    satisfies $\IC' \cup \dec$. %
    In turn, as $\IC'$ is over $R$, $I$ is also a model of $\IC'$. %
    By Lemma~\ref{lem:model-xuinds}, there exists an instance $J' \supseteq I$
    satisfying all of the X-UINDs in $\cl$ and, by Lemma~\ref{lem:model-yuinds},
    there exists $J'' \supseteq J'$ satisfying all of the Y-UINDs in $\cl$. %
    Moreover, by construction, for each tuple in $J''$ there is a tuple in $J'$
    having the same values at non-interpreted positions, thus $J''$ also
    satisfies all of the X-UINDs in $\cl$. %
    Therefore, $J''$ is model of $\cl$ and, as $\cl \equiv \IC$, of $\IC$ as
    well. %
    Let $J$ be the instance over $\R \cup \V$ with $R^J = R^{J''}$ (the
    extension of each $V_i$ under $J$ is unambiguously determined by $R^J$). %
    Clearly, $J \models \IC \cup \dec$ but $J \not\models \phitilde$, because
    $\overline{t} \in R^J$ while $\overline{t} \not\in {{V_1}^J \cup \dotsb \cup
      {V_n}^J}$.\hfill\IEEEQED
  \end{description}
\end{thmpf} 

Observe that Theorems~\ref{thm:yuind-sep} and~\ref{thm:xuind-sep} are direct
corollaries of Theorem~\ref{thm:uind-sep}.
The proof of Theorem~\ref{thm:yuind-sep} is analogous to the one above, with the
difference that in the ``only if'' direction, as $\IC$ does not contain X-UINDs,
Lemma~\ref{lem:model-xuinds} is not needed in order to build the instance $J'$
(simply take $J' = I$), and therefore the CDCs are not required to be globally
consistent.
The proof of Theorem~\ref{thm:xuind-sep} is also very similar to the one above,
with the difference that, since $\IC$ does not contain Y-UINDs, there is no need
to compute $\cl$, which in this case is always equal to $\IC$. %
Hence, we obtain $\emptyset$-separability rather than
$\{\eqref{eq:dom-prop}\}$-separability. %
The ``if'' direction works also with $\cl = \IC$, which is indeed a special
case, while for the ``only if'' direction one can simply take $J'' = J'$ (as
Lemma~\ref{lem:model-yuinds} is not needed).

Next, we show that replacing global consistency of the CDCs in the assumptions
of Theorem~\ref{thm:uind-sep} by disjointness w.r.t.\ the X-UINDs yields another
sufficient condition for the $\{\eqref{eq:dom-prop}\}$-separability of the UINDs
from the CDCs.

\begin{theorem}
  \label{thm:uind-sep-disj}
  Let $\IC$ be a set of CDCs and UINDs such that the CDCs are disjoint w.r.t.\
  each X-UIND in $\IC$, and the CDCs and Y-UINDs are dp-controllable. %
  Then, the UINDs are $\{\eqref{eq:dom-prop}\}$-separable in $\IC$ from the
  CDCs.\hfill\IEEEQED
\end{theorem} 

\noindent The proof of the above theorem is analogous to that of
Theorem~\ref{thm:uind-sep}, with the difference that in the ``only if''
direction the existence of the instance $J'$ is guaranteed by the following
lemma rather than Lemma~\ref{lem:model-xuinds}.

\begin{lemma}
  \label{lem:model-xuinds-disj}
  Let $\IC$ be a set of X-UINDs and CDCs such that the CDCs are disjoint w.r.t.\
  each X-UIND in $\IC$, and let $I$ be a model of $\cdc(\IC)$. %
  Then, there exists an instance $J$ such that $J \supseteq I$ and $J \models
  \IC$.
\end{lemma} 

\newcommand{\pos}{\operatorname{\sf pos}}

\begin{pf}
  The construction of $J$ is the same as in Lemma~\ref{lem:model-xuinds}, with
  the only difference that, in step 3 of the procedure, the existence of
  suitable values for $\overline{t}'$ at interpreted posi\-tions is guaranteed
  by the disjointness of the CDCs w.r.t.\ the X-UINDs, as shown below.

  We say that a CDC \emph{applies} on an instance, if that instance contains a
  tuple whose values at non-interpreted positions make the antecedent of the CDC
  true. %
  Denote by $\IC'_k$ and $\IC'_{k+1}$ the sets of CDCs in $\IC'$ that apply on
  $J_k$ and $J_{k+1}$, respectively. %
  For a CDC $\phi$, let $\pos(\phi)$ be the set of (interpreted) positions
  corresponding to the variables mentioned in the consequent of $\phi$.
  \begin{enumerate}[a), widest=b]
  \item Let $\psi \in \IC'_{k+1} \cap \IC'_k$. %
    Clearly, there exist suitable values for $\overline{t}'$ at interpreted
    positions so that $\{ R(\overline{t}') \} \models \psi$ (just take the
    corresponding values from $\overline{t}$).

  \item Let $\phi \in \IC'_{k+1} \setminus \IC'_k$. %
    The values at interpreted positions in $\overline{t}$ and $\overline{t}'$
    differ only at position $j$, thus the antecedent of $\phi$ mentions the
    variable $x_j$. %
    Then, since position $j$ is affected by the r.h.s.\ of the X-UIND $R[i]
    \subseteq R[j]$ under consideration, by Definition~\ref{def:disjoint-cdcs}
    the consequent of $\phi$ is satisfiable and the variables occurring therein
    are not mentioned in any other CDC in $\IC'$. %
    Thus, there exist suitable values for $\overline{t}'$ at interpreted
    positions such that $\{ R(\overline{t}') \} \models \phi$, where the values
    at interpreted positions not in $\pos(\phi)$ can be chosen freely.

  \item Let $\phi,\phi' \in \IC'_{k+1} \setminus \IC'_k$ and $\psi \in
    \IC'_{k+1} \cap \IC'_k$, then  $\pos(\phi)$, $\pos(\phi')$ and $\pos(\psi)$
    are pairwise disjoint.
  \end{enumerate}
  From all of the above, we conclude that there are suitable values for
  $\overline{t}'$ at interpreted positions so that the instance $\{
  R(\overline{t}') \}$ satisfies $\IC'_{k+1}$ and, in turn,
  $\IC'$.\hfill\IEEEQED
\end{pf}

Theorem~\ref{thm:xuind-sep-disj} is a direct consequence of
Theorem~\ref{thm:uind-sep-disj}, in the same way as Theorem~\ref{thm:xuind-sep}
follows from Theorem~\ref{thm:uind-sep}.

\subsection{FDs and UINDs Together}
\label{sec:uinds-fds}

Unfortunately, the separability results presented above for combinations of CDCs
and UINDs do not automatically carry over to the case in which FDs are also
present. %
In fact, although FDs do not directly interact with CDCs, they do in general
interact with UINDs,\footnote{The interaction between FDs and UINDs can be fully
  captured, as there is a sound and complete axiomatization for the finite
  implication of FDs and UINDs \cite{Abiteboul:1995:FD}.} which in turn interact
with CDCs.


We write FDs over $R$ as implications between sets of positions of $R$ (e.g.,
$\{1,3\} \rightarrow \{4\}$). %
We call \emph{X-FD} (resp., \emph{Y-FD}) an FD whose l.h.s.\ and r.h.s.\ both
consist of non-interpreted (resp., interpreted) positions; and we call
\emph{XY-FD} (resp., \emph{YX-FD}) an FD where the l.h.s.\ consists of
non-interpreted (resp., interpreted) positions and the r.h.s.\ of interpreted
(resp., non-interpreted) ones.

The following generalises Theorem~\ref{thm:yuind-sep} in the presence of X-FDs
and YX-FDs.

\begin{theorem}
  \label{thm:xfd-yuind-sep}
  Let $\IC$ be a set of CDCs, Y-UINDs, X-FDs and YX-FDs, where the CDCs and
  Y-UINDs are dp-controllable. %
  Then, the X-FDs, YX-FDs and Y-UINDs are $\{ \eqref{eq:dom-prop} \}$-separable
  in $\IC$ from the CDCs.
\end{theorem} 

\begin{pf}
  The proof given for Theorem~\ref{thm:uind-sep} can be modified as follows: in
  the ``only if'' direction take $J' = I$, which contains only the tuple
  $\overline{t}$, and construct $J''$ as in Lemma~\ref{lem:model-yuinds} (with
  $\IC^* = \IC$) by extending $J'$ with tuples that have the same values as
  $\overline{t}$ at non-interpreted positions. %
  Therefore, $J''$ satisfies any FD whose r.h.s.\ is a set of non-interpreted
  positions.\hfill\IEEEQED
\end{pf}

Theorem~\ref{thm:xuind-sep} does not hold anymore in the presence of Y-FDs, that
is, X-UINDs and Y-FDs are not $\emptyset$-separable in general from globally
consistent CDCs, as shown below.

\begin{theorem}
  \label{thm:yfd-xuind-notsep-gc}
  There is a set of X-UINDs, Y-FDs and globally consistent CDCs, in which the
  X-UINDs and Y-FDs are not $\emptyset$-separable from the CDCs.
\end{theorem} 

\begin{pf}
  Let $R$ be a relation symbol of arity $4$, whose last two positions are
  interpreted over the integers. %
  Let $\IC$ consist of the X-UIND $R[1] \subseteq$ $R[2]$, the Y-FD $R \colon
  \{3\} \rightarrow \{4\}$, and the following CDCs:
  \begin{align*}
    x_1 = a \land x_2 = b &\,\rightarrow\, y_1 = 0 \land y_2 > 1 \enspace ;\\
    x_2 = a &\,\rightarrow\, y_1 = 0 \land y_2 < 1 \enspace .
  \end{align*}
  The above CDCs are globally consistent, since their consequents are
  satisfiable and their antecedents are never true at the same time (as $x_2$
  cannot be simultaneously equal to $b$ and $a$).
  Let $\dec$ be the horizontal decomposition specified by $V_1 \colon x_1 \neq
  a$ and $V_2 \colon x_2 \neq b$. %
  Clearly, $\dec$ is lossy under $\cdc(\IC)$ as the instance $I = \{ R(a,b,0,2)
  \}$ satisfies $\cdc(\IC)$ and $\dec$; indeed, the tuple $(a,b,0,2)$ is in
  $R^I$ but it is not selected by either $V_1$ or $V_2$.
  Suppose that $\dec$ is lossy under $\IC$. %
  Then, there exists a model $J$ of $\IC \cup \dec$ and a tuple $\overline{t}
  \in R^J$ such that $\overline{t} \not\in {V_1}^J \cup {V_2}^J$. %
  By definition of $V_1$ and $V_2$, we have that $\overline{t}[1] = a$ and
  $\overline{t}[2] = b$ and, in turn, $\overline{t}[3] = 0$ and $\overline{t}[4]
  > 1$ by the first CDC. %
  By the X-UIND, there must be $\overline{t}' \in R^J$ such that
  $\overline{t}'[2] = a$ and, in turn, $\overline{t}'[3] = 0$ and
  $\overline{t}'[4] < 1$ by the second CDC. %
  But then, $\overline{t}$ and $\overline{t}'$ violate the Y-FD, since they
  agree on the third position but must differ on the fourth. %
  Hence, $J \not\models \IC$, which is a contradiction. %
  Therefore, $\dec$ is lossless under $\IC$, and we conclude that the X-UIND and
  Y-FD are not $\emptyset$-separable in $\IC$ from the CDCs.\hfill\IEEEQED
\end{pf}

The CDCs in the above proof are globally consistent, but not disjoint w.r.t.\
the X-UIND. %
However, Theorem~\ref{thm:xuind-sep-disj} does not hold either in the presence
of Y-FDs, that is, not even disjointness is enough to ensure the
$\emptyset$-separability of X-UINDs and Y-FDs from CDCs.

\begin{theorem}
  \label{thm:yfd-xuind-notsep-dis}
  There exists a set of CDCs, X-UINDs and Y-FDs, in which the CDCs are disjoint
  w.r.t.\ each X-UIND, but the X-UINDs and Y-FDs are not $\emptyset$-separable
  from the CDCs.
\end{theorem} 

\begin{pf}
  Let $R$ be a relation symbol of arity $4$ with its last two positions
  interpreted over the integers. %
  Let $\IC$ consist of the X-UIND $R[1] \subseteq$ $R[2]$, the Y-FD $R \colon
  \{3\} \rightarrow \{4\}$, and the CDC $x_2 = a \rightarrow y_1 = 0 \land y_2 =
  2$, trivially disjoint with the X-UIND.
  Consider the horizontal decomposition $\dec$ specified by $V \colon x_1 \neq a
  \land x_2 \neq b \land y_1 \neq 0 \land y_2 \neq 1$. %
  Clearly, $\dec$ is lossy under $\cdc(\IC)$ because the instance $I = \{
  R(\overline{t}) \}$, where $\overline{t} = (a,b,0,1)$, satisfies $\cdc(\IC)$
  and $\dec$; indeed, $\overline{t}$ is not selected by $V$.
  Suppose that $\dec$ is lossy under $\IC$; since $V$ selects any tuple other
  than $\overline{t}$, there is a model $J$ of $\IC \cup \dec$ such that
  $\overline{t} \in R^J$ but $\overline{t} \not\in V^J$. %
  By the X-UIND, there must be $\overline{t}' \in R^J$ such that
  $\overline{t}'[2] = a$ and, in turn, $\overline{t}'[3] = 0$ and
  $\overline{t}'[4] = 2$ by the CDC. %
  But then, $\overline{t}$ and $\overline{t}'$ violate the Y-FD, because they
  agree on the third position but differ on the fourth. %
  So $J \not\models \IC$, which is a contradiction. %
  Hence, $\dec$ is lossless under $\IC$, and we conclude that the X-UIND and
  Y-FD are not $\emptyset$-separable in $\IC$ from the CDCs.\hfill\IEEEQED
\end{pf}



\section{Discussion and Outlook}
\label{sec:conclusion}

In this article, we studied lossless horizontal decomposition under constraints
in a setting where the values for some of the attributes in the schema are taken
from an \emph{interpreted} domain. %
Data values in such a domain can be compared in ways beyond equality, according
to a first-order language $\lang{C}$. %
We did not make any assumption on $\lang{C}$, other than requiring it to be
closed under negation.

In the above setting, we considered a class of integrity constraints, CDCs,
based on those introduced in~\cite{FeinererFG:2013:bncod}.
We have characterised the consistency of a set of CDCs in terms of
satisfiability in $\lang{C}$ and we have shown that the problem of deciding
consistency is \cplex{NP}-complete when \lang{C} is the language of either
UTVPIs or BUTVPIs.

We considered a more general form of selections than in
\cite{FeinererFG:2013:bncod} and characterised, in terms of unsatisfiability in
$\lang{C}$, whether a horizontal decomposition specified by such selections is
lossless under CDCs. %
We have shown that the problem of deciding losslessness is
\cplex{co-NP}-complete when \lang{C} is the language of either UTVPIs or
BUTVPIs.

We also considered losslessness under CDCs in combination with FDs and UINDs. %
We introduced and studied the important notion of separability, which indicates
whether constraints other than CDCs can be disregarded w.r.t.\ losslessness,
after incorporating the effect of their interaction in terms of entailed CDCs. %
A summary of all the separability results presented in this article is given in
Table~\ref{tab:summary}.

\begin{table}
  \caption{\small Summary of $\rules{S}$-separability results
    (unr~=~unrestricted, dpc~=~dp-controllable,
    gc~=~globally~consistent, dis~=~disjoint).}
  \label{tab:summary}
  \centering
  \begin{tabularx}{\linewidth}{Xccc}
    \toprule
    \textbf{Constraints} & \textbf{CDCs} & $\rules{S}$ & \textbf{Theorem}
    \\
    \midrule
    FDs & unr & $\emptyset$ & \ref{thm:adding-fds}
    \\
    \midrule
    Y-UINDs & \multirow{2}{*}{dpc} & \multirow{2}{*}{$\{\eqref{eq:dom-prop}\}$}
    & \protect\ref{thm:yuind-sep}
    \\
    (+ X-FDs + YX-FDs) & & & \protect\ref{thm:xfd-yuind-sep}
    \\
    \midrule
    \multirow{2}{*}{X-UINDs} & gc & \multirow{2}{*}{$\emptyset$} &
    \protect\ref{thm:xuind-sep}
    \\
    & dis & & \protect\ref{thm:xuind-sep-disj}
    \\
    \midrule
    \multirow{2}{*}{X-UINDs + Y-UINDs} & 
    \parbox{\widthof{dpc + dis}}{dpc + gc} &
    \multirow{2}{*}{$\{\eqref{eq:dom-prop}\}$} & \protect\ref{thm:uind-sep}
    \\
    & dpc + dis & & \protect\ref{thm:uind-sep-disj}
    \\
    \bottomrule
  \end{tabularx}
\end{table} 

A promising direction for future research we are currently investigating is the
generalisation of the separability results for UINDs to arbitrary inclusion
dependencies (INDs). %
Observe that INDs, differently from UINDs, can affect both interpreted and
non-interpreted attributes at the same time, e.g., in $R[x_1,y_1] \subseteq
R[x_2,y_2]$. %
Some care is needed in allowing FDs in this setting as well, because log\-ical
implication for unrestricted combinations of FDs and INDs is undecidable and has
no axiomatization \cite{Abiteboul:1995:FD}. %

Another interesting direction is that of allowing equalities between two
variables in the antecedents of CDCs as well as in the selection conditions on
non-interpreted attributes of view definitions. %
We believe our approach could be extended in this direction by representing such
equalities by propositional variables and by adding suitable axioms to the
auxiliary theory to handle transitivity and symmetry.

The main motivation for our study of lossless horizontal decomposition is that
it provides the groundwork for the consistent and unambiguous propagation of
updates in the context of selection views. %
By applying the general criterion of \cite{Franconi:2012:TVU}, given a lossless
horizontal decomposition it is possible to determine whether an update issued on
some (possibly all) of the fragments can be propagated to the underlying
database without affecting the other fragments. %
Similarly, it is possible to partition the source relation by adding suitable
conditions in the selections that define the fragments, so that each is disjoint
with the others. %
In general, a lossy horizontal decomposition can always be turned into a
lossless one by defining an additional fragment, called a \emph{complement},
which selects the missing tuples. %
In particular, there is a \emph{unique minimal complement} selecting all and
only the rows of the source relation that are not selected by any of the other
fragments. %
In follow-up work, we will show how to compute the definition of such a
complement, in the scope of an in-depth study of partitioning and update
propagation in the setting studied in this article.


Most of the work in the field of horizontal decomposition has been carried out
in the context of distributed databases systems, where one is mainly concerned
with finding an optimal decomposition w.r.t.\ some parameters (e.g., workload,
query-execution time, storage quotas), rather than determining whether a
\emph{given} horizontal decomposition is lossless.

De Bra (\cite{DeBra:1987:hdr,DeBra:1986:icdt}) developed a theory of horizontal
decomposition to partition a relation into two sub-relations such that one
satisfies certain FDs that the other does not. %
The approach is based on constraints that capture partial implications between
sets of FDs and exceptions to sets of FDs, for which a sound and complete set of
inference rules is provided. %
These constraints are $\emptyset$-separable from our CDCs (for the same reason
FDs are).

Maier and Ullman \cite{MaierU:1983:sigmod} consider horizontal decomposition
involving physical and virtual fragments over the same attributes. %
Fragments are defined in an arbitrary (first-order) language closed under
Booleans, where entailment is decidable and consisting of formulae that, as in
our case, can be evaluated by examining one tuple at a time, in isolation from
the others. %
Differently from our case, the language allows to express equalities between
variables associated with non-interpreted attributes. %
But, if such equalities are forbidden, the setting of \cite{MaierU:1983:sigmod}
can be recast into ours: the union of the physical fragments is the single
source relation $R^I$ we consider here, the definitions of the physical
fragments can be taken as integrity constraints over $R$, and the definition of
each virtual fragment (given in terms of the physical fragments and other
virtual ones) can be expressed only in terms of $R$ by query unfolding.
%
%
Then, the problem of determining whether the virtual fragments constitute a
lossless horizontal decomposition of the physical fragments, which is not
addressed in \cite{MaierU:1983:sigmod}, can be solved by applying the techniques
we described in this article. %
Virtual fragments in \cite{MaierU:1983:sigmod} are defined by selection and
union, that is, in our notation, by formulae of either the form
$\cond(\overline{x}) \land \sel(\overline{y})$ or
$\cond(\overline{x}) \lor \sel(\overline{y})$. %
As we remarked in Section~\ref{sec:prelim}, in such a case losslessness can be
checked by considering two views $\cond(\overline{x})$ and $\sel(\overline{y})$
in place of each view of the latter form.

\ifCLASSOPTIONcompsoc
  \section*{Acknowledgements}
\else
  \section*{Acknowledgement}
\fi

The authors would like to thank the anonymous review\-ers for their valuable
comments and suggestions, which contributed to improving the quality of the
paper. %
They also thank Jef Wijsen for pointing out errors in~\cite{Feinerer:2015:LSV}
(cf.\ Definition~\ref{def:disjoint-cdcs} and
Example~\ref{exa:incomp-glob-disj}).

\ifCLASSOPTIONcaptionsoff
\fi



\bibliographystyle{IEEEtran}
\bibliography{references}

%





\newlength{\biosep}
\setlength{\biosep}{-6mm}
\vfill
\vspace{\biosep}
\begin{IEEEbiography}[{\includegraphics[%
    width=1in, height=1.25in, clip, keepaspectratio]{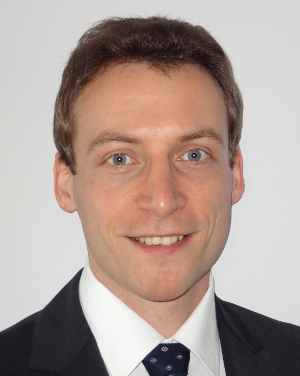}}]{%
    Ingo Feinerer}
  has been assistant professor at the Vienna University of Technology since
  2008. %
  He holds a PhD in computer science from the Vienna University of Technology
  and a PhD in business administration from the Vienna University of Economics
  and Business. %
  In 2007 he was postdoctoral visiting scholar at the computer science
  department of Carnegie Mellon University. %
  His expertise and research interests comprise configuration, formal methods,
  text mining, and databases.
\end{IEEEbiography}
\vspace{\biosep}
\begin{IEEEbiography}[{\includegraphics[%
    width=1in, height=1.25in, clip, keepaspectratio]{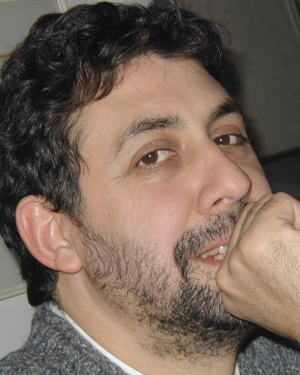}}]{%
    Enrico Franconi}
  is associate professor at the Free University of Bozen-Bolzano, Italy, where
  he is the director of the KRDB Research Centre for Knowledge and Data. %
  His main research interest is on knowledge representation and reasoning
  technologies applied to databases, and in particular description logics,
  temporal representations, conceptual modelling, intelligent access to
  information, natural language interfaces. %
  Recently he has been involved as principal investigator in many European
  projects (SEWASIE, DWQ), networks of excellence (KnowledgeWeb, InterOp), and
  Marie Curie actions (NET2). %
  He is a member of the Advisory Committee of the World Wide Web Consortium
  (W3C). %
  He is in the editorial board of the Journal of Applied Logic (JAL), published
  by Elsevier. %
  He is the author of more than 150 publications in international workshops,
  conferences, and journals.
\end{IEEEbiography}
\vspace{\biosep}
\begin{IEEEbiography}[{\includegraphics[%
    width=1in, height=1.25in, clip, keepaspectratio]{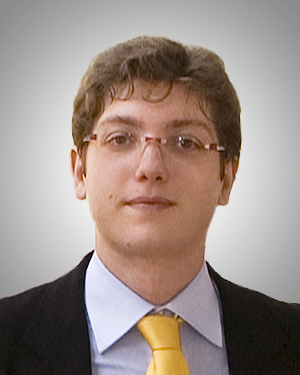}}]{%
    Paolo Guagliardo}
  completed the European Master in Computational Logic in 2009, obtaining the
  MSc degree in computer science with honours from the Vienna University of
  Technology and the Free University of Bozen-Bolzano. %
  He is currently a PhD candidate at the latter university under the supervision
  of Enrico Franconi. %
  His research interests include database theory, knowledge representation and
  semantic web.
\end{IEEEbiography}




\end{document}